# Surface spin canting in $Fe_3O_4$ and $CoFe_2O_4$ nanoparticles probed by high resolution electron energy loss spectroscopy

D. S. Negi,[1,2] H. Sharona,[1,2] U. Bhat,[1,2] S. Palchoudhury,[4] A. Gupta[3] and R. Datta[1,2]*

[1]*International Centre for Materials Science, Jawaharlal Nehru Centre for Advanced Scientific Research, Bangalore 560064, India.*

[2]*Chemistry and Physics of Materials Unit, Jawaharlal Nehru Centre for Advanced Scientific Research, Bangalore 560064, India.*

[3]*Center for Materials for Information Technology, University of Alabama, Tuscaloosa, Alabama 35487, USA.*

[4]*Civil & Chemical Engineering, University of Tennessee at Chattanooga, Chattanooga, Tennessee 37403, USA.*

*Corresponding author E-mail: ranjan@jncasr.ac.in

High resolution electron energy loss spectroscopy (HR-EELS) is utilized to probe the surface spin canting in nanoparticles of two technologically important magnetic materials, i.e. $Fe_3O_4$ and $CoFe_2O_4$ (CFO). A soft experimental technique is developed that is capable of extracting EELS spectra with one atomic plane resolution recorded in a single frame. This yields information at different depth of the nanoparticle from the surface to the core regions with high signal to noise ratio and without beam damage. This enables comparing the fine structures between the surface and core regions of the nanoparticles. The results confirm earlier observations of uniformly oriented spin canting structure for CFO with additional information on atom site-selective spin canting information. In case of $Fe_3O_4$ preferred canting orientation forming core and shell structure is deduced. Unlike earlier reports based on polarized spin-flip neutron scattering measurement, it is possible to narrow down the possible canting angles for $Fe_3O_4$ ($T_d$, $O_h$ tilts 40°, 40°) and CFO ($T_d$, $O_h$ tilts 17°, 17°) from the experimental spectra combined with the first principle based calculation considering non-collinear magnetism. In addition, the role of Dzyaloshinskii-Moriya interaction in stabilizing the spin canting at the nanoparticle surface is discussed. The results demonstrate that HREELS can be a powerful technique to probe the magnetic structure in nano-dimensional systems and has advantages over neutron based techniques in terms of superior spatial resolution, site specific information and easy of sample preparation.

I. **Introduction**

Magnetic nanoparticles particularly $Fe_3O_4$ and $CoFe_2O_4$ (CFO) have found wide range of applications in areas such as biomedicine for hyperthermia based cancer treatment, drug delivery, MRI contrast agents, bio-imaging, spintronics, high density data storage, etc. [1-10]. The prerequisite for all these applications is large magnetic response even at nanometer dimension. However, it is often found that the saturation magnetization in nanoparticles is significantly reduced compared to their bulk counterpart [11-13]. Surface spin canting due to broken symmetry, surface spin disordering, crystallographic changes and magnetic dead layers at the surface are generally considered to be responsible for the reduced saturation magnetization ($M_s$) with respect to their bulk counterpart. This significantly reduces the practical efficiency and sensitivity of such nanoparticles [14-18]. There is considerable interest in characterizing and understanding the surface spin structure of magnetic nanoparticles from the point of view of fundamental science as well as improvements in synthesis procedures [19-25]. Research is actively being pursued to understand the surface spin geometry for various nano-dimensional systems and a number of novel experimental techniques have been developed in the recent past to attain such information [26-36, 43].

Among various techniques to probe the surface spin geometry, 2D polarization analyzed small angle spin-flip neutron scattering (PASANS) has recently been developed to determine the three dimensional spatial distribution of spin moments in a dense face centered cubic assembly of iron oxide (Fe$_3$O$_4$, size ~ 9 nm) and CFO (CoFe$_2$O$_4$, size ~ 11 nm) nanoparticles [37, 38]. The method is based on first detecting a negative cross term in the neutron spin-flip scattered intensity ($I_{SF}(\vec{Q})$) and then fitting an energy balance model to estimate the possible sets of canting angles and shell thickness. A correlated phase factor $\overline{cos}(\delta\varphi)$ between magnetic distributions parallel ($|M_{\parallel\vec{H}}(\vec{Q})|$) and perpendicular ($|M_{\perp\vec{H}}(\vec{Q})|$) to the applied field indicates the presence of core-shell geometry in the ensemble of nanoparticles. The analysis revealed magnetic core-shell morphology and a uniformly canted structure for Fe$_3$O$_4$ and CFO nanoparticles, respectively. The energy balance model is dominated by Zeeman energy vs. exchange energy and Zeeman energy vs. anisotropy energy for Fe$_3$O$_4$ and CFO nanoparticles, respectively. The analysis inferred a range of canting angles $\varepsilon = 27°$ to $42°$ at 200 K and $\varepsilon = 23°$ to $31°$ at 300 K with corresponding shell thickness of 1.0 nm $\pm$ 0.2nm and 1.5nm $\pm$ 0.2 nm for Fe$_3$O$_4$. At 10 K, canting angles $\varepsilon$ in terms of $T_d$ tilt in the range of $50°$ to $85°$ with a wide mix of shell thicknesses is obtained. At 300 K and 0.005 T remnant field a preferred $T_d$ tilt = $5°$ with no definite shell thickness is derived based on the energy balance model. For CFO, canting angles of $33°$ and $17°$ have been deduced at 10 and 300 K, respectively. The effect of temperature has been introduced in the energy balance model in terms of $m \propto \beta$, where $\beta$ is the ratio of NP magnetization ($m$) to that of the bulk crystal ($m_s$) of the bulk crystal.

Though the technique is certainly a powerful and pioneering development in understanding magnetic geometry at the nanoscale, it requires a large assembly of nanoparticles forming an ordered crystal lattice to obtain such information, wherein presence of capping layer and inter-particle dipolar interaction cannot be avoided. Moreover, it may often be difficult to form such an ordered crystal lattice of the nanoparticles with internal crystallographic symmetry aligned between them, which will likely lead to significant scatter in the recorded spin distributions. Moreover, smaller particle size may give rise to additional problems.

Alternative experimentation has been conducted in a transmission electron microscope (TEM) on individual magnetic nanoparticles based on electron energy loss spectroscopy (EELS) to obtain equivalent magnetic information with high spatial and energy resolution. Two different EELS based techniques have been utilized so far to investigate the surface magnetism in nanoparticles; electron magnetic chiral dichroism (EMCD) and spatially resolved high resolution electron energy loss spectroscopy (HR-EELS) [39, 26]. The principle of EELS is based on interpreting the signal in terms of density of unoccupied states which is sensitive to any changes in the electronic structure in the material and in the present case it is the geometrical arrangement of spins. While EMCD revealed qualitative difference (~30%) in magnetic order between the surface and interior of $Fe_3O_4$ nanoparticles [39], high resolution EELS (HR-EELS) technique has been able to quantify the difference in overall magnetic order between the regions with and without capping agents in the case of $CuCr_2S_4$ nanoparticles [26]. The HREELS based technique indicated that capping agents help significantly restore the magnetic moment at the surface of the nanoparticles. Moreover, it helps to explain the large difference in $m_s$ values between the spectroscopy and bulk magnetometry techniques in terms of unaccounted weight of the capping agents, which significantly underestimates the magnetization of nanoparticles by bulk magnetometry techniques [11, 13, 26].

In the present report, we have further extended the HREELS technique to probe the surface spin canting in $Fe_3O_4$ and CFO nanoparticles. The technique is based on experimentally recording the fine features in the HREELS spectra combined with the first principle-based calculation using WIEN-NCM code [40]. The code can simulate changes in electronic structure due to spin canting at various angular configurations with respect to the usual ferrimagnetic configuration. Additionally, we have considered the Dzyaloshinskii-Moriya (or DM) interaction due to spin canting and ascertained that DM energy savings can also stabilize the spin canting geometry at the surface due to broken symmetry other than Zeeman energy term, which otherwise requires application of an external field. However, the Zeeman term is important since it further helps form a core-shell structure. In order to understand the surface spin structure and spatial distribution of magnetic response from a single nanoparticle it is essential to distinguish the magnetic response from the core and surface regions separately. Therefore, the experimentation required developing a soft technique in order to obtain high quality spectra capable of providing single atomic plane resolution without damaging the particles, along with all the information being recorded in a single exposure frame. The usual choice of STEM-EELS combination usually results in drilling holes in most nanoparticle samples, thus preventing the collection of spectra for sufficiently long exposure time [26, 41]. The overall results of this study are consistent with those from previous neutron-based experiment suggesting that a core and canted shell is formed in the case of $Fe_3O_4$ and uniformly canted configuration occurs in the case of CFO. Additional information is obtained on atom site-selective spin canting. Moreover, the first-principle-based method in combination with the experimentation have helped to narrow down the canting angle for $Fe_3O_4$ ($T_d$, $O_h$ tilts 40°, 40°) and CFO ($T_d$, $O_h$ tilts 17°, 17°) for the measurements carried out at 300 K. We infer different shell thicknesses at two different temperatures, i.e. 77 K and 300 K, for $Fe_3O_4$ and for CFO. The results indicate that HR-EELS can indeed be used to probe the fine details of spin structure at the nanoparticles surface and the same can be extended to other nano-dimensional magnetic systems.

II. **Experimental and Theoretical Techniques**

Nanoparticles of $Fe_3O_4$ and CFO were prepared by thermolysis of $Fe^{+3}$-oleate and mixed $Co^{+2}$ $Fe^{+3}$-oleate complex, respectively, following our previously published protocol. [13, 42]. In a typical synthesis of $Fe_3O_4$ nanoparticles, iron oleate (2 mmol) is thermally decomposed at 320 °C for 2.5 h in the presence of oleic acid (0.1 mL)/tri-octylphosphine oxide (0.2 g) surfactant mixture in 1-octadecene under a $N_2$ atmosphere. Similarly, mixed $Co^{2+}Fe^{3+}$-oleate precursor (2 mmol) is heated at 320 °C for 1 h under inert gas protection in the presence of capping agent, oleic acid (0.2 g) dissolved in 1-octadecene (6 mL), to form CFO nanoparticles.

High resolution transmission electron microscopy (HRTEM) and high resolution electron energy loss spectroscopy (HR-EELS) are performed in a double aberration correction transmission electron microscope FEI TITAN$^{3TM}$ 80-300 kV equipped with a gun monochromator. All the spectra are collected with a GIF entrance aperture of 1 mm and energy dispersion of 0.03 eV/channel. For achieving high spatial resolution, the obvious choice is to work with a STEM-EELS combination mode. However, for most of the crystals the STEM probe drills holes in the area of interest [Fig. S1 (a)-(c)] due to high beam current before a quality spectrum can be recorded. Therefore, we have developed a soft experimental technique based on EELS with monoprobe [Fig. 1(a)], where first the nanoparticle of interest is placed at the center of the GIF entrance aperture [Fig. 1(b)] with atomic resolution image and then collecting the spectra in $Y$ vs. $\Delta E$ equivalent to $q$ vs. $\Delta E$ in the diffraction image [Fig. 1(c)]. The images and spectra have direct correlation in terms of spatial information and corresponding spectra when one of the two spatial dimensions is folded or projected at every point to the other perpendicular axis. The spectra extracted from each slice [Fig. 1(c)] have information from the projected area of the nanoparticles as marked in the Figure 1(b). Extracted spectra are shown in Figure 1(d) & (e). With this method it is possible to obtain spatial resolution of one atom plane; however, the narrowest slice width chosen is ~ 1-2 nm, which is sufficient for the present investigation. The added advantage of the technique is that it permits to collect spectra in low dose parallel illumination mode and thus allows acquisition of spectra for sufficiently long exposure time without damaging the specimens along with monoprobe illumination, which is essential to preserve high energy resolution information. Moreover, all the spatial information is encoded in one single acquisition frame. Previously, a similar approach has been utilized, except for very high spatial resolution where a probe area as small as ~2 sq. nm is obtained by magnifying the specimen and using GIF as a selected area aperture and any loss in signal due to magnification can be compensated by de-magnifying the mono-probe on the sample, which provides an independent control system in a microscope equipped with a gun monochromator [26]. The experiments have been performed at two different temperatures namely, 77 K and 300 K. The described method can be extended to extract EMCD signal from a spatially resolved nanometer length scale area [Ref. 43, Fig. S2, S3].

We have carried out first principle based calculation of total energy, cohesive energy, density of states and EELS spectra for both $Fe_3O_4$ and CFO using WIEN2k code [44]. WIEN2K is a full potential LAPW + LO method within the framework of density functional theory. Various canting configurations of tetrahedral and octahedral tilts, i.e. $T_d$, $O_h$ as well as relative tilts between them such as both tilt along reference axis and azimuthal orientations are considered for calculating relative stability over typical ferrimagnetic configurations using magnetic non-collinear WIEN-NCM code with atomic moment approximation (AMA) [40]. Figure 2 shows the reference axis for various tilt and azimuthal directions of spins with respect to the ferrimagnetic alignment. A schematic of various canted structures can be found in Ref. 43. The lattice parameters are optimized with Perdew–Burke–Ernzerhof (PBE) functional within the generalized gradient approximation (GGA). The $RK_{max}$ value is set to 7.0. Further energy correction is done with GGA+U method, with U, J values taken from the Ref. 45, 46. The magnetic moment of $Fe_3O_4$ and CFO unit cell are 4 and 3 µB, respectively that is consistent with previous reports [47, 48].

### III. Results and Discussion

Figure 3 (a), (b) & (c), (d) show the low magnification and HRTEM images of well dispersed nanoparticles of $Fe_3O_4$ and CFO, respectively. Most of the particles are single crystalline and the crystallinity is preserved even at the surface. Particles with good crystallinity, i.e. without any visible crystallographic defects with distinct shape and size are selected for the present investigation, as shown in the example individual particle image. The particles are capped with oleic acid for both $Fe_3O_4$ and CFO, which ensures dispersion and prevents agglomeration. The overall morphology of the particles is spherical with average diameters of 25 (±2) nm and 10 (±4) nm for Fe$_3$O$_4$ and CFO nanoparticles, respectively [13].

### A. Fe$_3$O$_4$ nanoparticles

Figure 4 (a) & (b) show representative experimental spectra for the Fe-$L_3$ edge of Fe$_3$O$_4$ nanoparticle from the core and interior regions at both 300 and 77 K. The spectra have been extracted by using a rectangular slice tool from the $Y$ vs. $\Delta E$ plot as describe in the experimental section [Fig. 1 (c)]. The slice width of the rectangular box is approximately 1-2 nm and 5-10 nm for the locations at the edge and center regions of the nanoparticles, respectively. This amplifies the relative spectral weights from the surface and bulk regions, respectively. Kindly see the discussion later in the section III (B) on how varying the rectangular slice width helps to estimate the approximate shell thickness of the nanoparticles. We have considered features only in the $L_3$ absorption edge and the complete spectra i.e. $L_{3,2}$ is given in Figure 1(d)&(e). The most significant tetrahedral and octahedral DOS contributions to the overall spectra are marked with arrows. The blue and red color arrows indicate the tetrahedral and octahedral site contributions to the spectra, respectively. This has been done with the help of theoretically simulated spectra [Fig. 5(a)]. The changes in features due to surface spin canting are marked in the spectra with black arrows. In order to understand the changes in the features due to spin canting geometry it is important to compare the results with the unoccupied DOS calculated by first principle method, which serves as a finger print in the absence of a standard experimental spectra with known spin canting configurations. As already mentioned, the EELS spectra contain information on the density of unoccupied states and are expected to be sensitive to the changes in electronic structure of the materials due to various spin canting arrangements. The changes are small and need careful analysis to discern the effect [49].

Figure 5(a) shows the simulated theoretical contributions from the tetrahedral and octahedral Fe to the overall EELS spectra. The partial $d$ orbital contributions for the respective tetrahedral and octahedral atom is shown in Figure 5 (b)&(c). The evolution of DOS has been studied systematically with different canting configurations. Various canting geometries have been considered and relative energy difference between them is given in Ref. 43. From the set of canting configurations considered for the calculation, the most stable configuration is $T_d$, $O_h$ tilt of 40°, 40° and the least stable is 20° (Table 1 (a), Ref. 43). Various azimuthal angles of spin orientations have also been considered for the most stable configurations and found not to change the stability of the system significantly (of the order of ~ 0.02 meV) [Table 1(b)]. The most stable canting configuration obtained from the present theoretical calculation falls in the higher side of the tilt range determined previously [37]. However, depending on the Zeeman energy term and competing energy, i.e. exchange energy in case of $Fe_3O_4$ might prefer canting angle with higher net magnetic moment for the shell. But, the difference in magnetic moment between comparable canting geometries is not sufficiently different so that Zeeman energy contribution under a high enough magnetic field can supersede the overall penalty energy term of the system and thus assume the canting configurations with the higher magnetic moment. The density of unoccupied DOS for comparable canting configurations including the most stable $T_d$, $O_h$ tilt of 40°, 40° with respect to ferrimagnetic configuration is shown in Figure 6 for both tetra and octahedral atomic sites. The primary difference between the ferrimagnetic to the canting cases is a decrease in band width of neighboring DOS along with an increase in the energy gap between the DOS peaks with more discrete nature in case of tetrahedral sites. For octahedral atomic sites, readjustment in the energy position of DOS and an increase in gap between the lower to the higher energy DOS can be noticed. However, it is difficult to distinguish features between the various types of canting configurations except for slight differences in relative peak heights and this require careful correlation-based analysis and can be the prospect of a future work. Therefore, for the present investigation and considering the very first attempt to utilize HREELS to probe canting geometry we have limited ourselves to consider only the changes in the separation of peaks and their discrete nature to identify the occurrence of canting at the surface of the nanoparticles and roughly estimate the canting shell thickness. Therefore, the difference in the DOS between un-canted and canted states can be understood in terms of relative changes in the orbital overlap thus leading to changes in terms of the discrete nature and relative spacing between the DOS peaks.

For the spectra recorded at 77 K, additional peaks are observed which are marked in Figure 4(b). This is similar to the 300 K spectra except for more number of peaks. These additional peaks are due to the monoclinic structure of $Fe_3O_4$ (space group 9 Cc) and is a result of well-known Verwey transition at temperatures below 120 K in this system. The calculation for this monoclinic structure also shows presence of peaks which are different in number and positions in energy scale contributing to the overall spectra [Ref. 43 (Fig. S7)]. Though discrete nature from the surface areas indicate presence of canting; however, it will be worth performing first principle calculation for various spin canting configurations for such monoclinic structure in order to determine the optimum spin canting angle in comparison to the inverse spinel structure. Similar difference in spectra at two different temperatures is not observed in case of CFO [see section B] demonstrating that the HR-EELS is capable of detecting signatures due to such changes in the structural symmetry.

The primary difference between the energy balance model used in past reports and the present first principle based calculation is that for the former, analysis of the various energy cost associated with canting has been evaluated from the component of canted spin moment magnitude through the $[1 - \cos(\varepsilon)]$ dependence term, where $\varepsilon$ is the canting angle at either the $T_d$ or $O_h$ site. Whereas, the first principle methods are quantum mechanical calculations and considers relative exchange and anisotropy energy cost for various noncollinear spin canting orientations with respect to other collinear spins in the crystal lattice. In fact, spin canting may increase or decrease the net magnetic moment in the unit cell [Table 1, Ref. 43] that contributes to the Zeeman energy savings or cost under an applied magnetic field that balances the dominant exchange energy and anisotropy energy cost in case of $Fe_3O_4$ and CFO, respectively. Spin canting is also associated with a noncollinear relationship in terms of spin arrangement with respect to collinear spins and will have additional DM interaction term in the general form of bilinear coupling energy between two spins [50, 51]. This noncollinear relationship between spin vectors in the crystal is responsible for the existence of DM term in case of $Sr_2IrO_4$ but cancellation in case of $Ba_2IrO_4$ [52, 53]. However, DM interaction has thus far not been considered in the context of surface spin canting of magnetic nanoparticles. The anisotropic exchange energy due to DM interaction can also contribute to the stability of the canted spin configurations at the nanoparticle surface due to reduced symmetry in addition to the Zeeman energy term, which appears in the presence of an external magnetic field. Therefore, additionally we have explored in the present report the extent to which the DM term plays a role in stabilizing spin canting at the nanoparticle surface and balancing the dominant counter energy term. DM interaction is generally used to the observed weak ferromagnetism in antiferromagnetic materials where the interaction strength is either few percent or equivalent to the isotropic energy term [54]. Recently, methods have been developed to study the anisotropic magnetic coupling related to the DM coupling parameter that involves first mapping magnetically constrained noncollinear DFT onto a general spin Hamiltonian [52, 55-58], then by fitting the spin related penalty contributions to the total energy, relative contributions and balance between Heisenberg exchange, the DM interaction and the single ion anisotropy terms can be evaluated. The penalty energy can be written for the spin constrained calculation as [53]

$$\Delta E = E - E_0 = \sum_i \gamma \ [M_i - M_i^0(M_i.M_i^0)]^2$$

Where $E_0$ is the DFT energy and the $\Delta E$ is the penalty energy due to noncollinear directional constraint. $M_i^0$ is a unit vector along the global direction of the magnetic moment at site $i$ and $M_i$ is the integrated magnetic moment inside the Wigner-Seitz cell around atom $i$, and $\gamma$ is the parameter through which the penalty energy term is controlled.

By, varying the spin canting angle we obtain DFT+SOC+U total energy plot as shown in Fig. 7. Two different plots are shown, the first one is with varying canting angle of $T_d$ site and the second one with varying the azimuthal angle for the most stable $T_d$ canting angle (40°). However, we have fitted the competing energy terms only for the first case as energy difference between different azimuthal angles is an order of magnitude smaller. The penalty energy is then mapped onto the standard expression of the classical spin Hamiltonian [53]

$$\Delta E = -\sum_{i<j} J_{ij} S_i \cdot S_j + \sum_i \varepsilon_{an}^i(S_i) + \sum_{i<j} D_{ij} \cdot [S_i \times S_j] \qquad (1)$$

Where, the first, second, and third terms represent the isotropic Heisenberg exchange, the single ion anisotropy, and DM interaction. The above equation for ferrimagnetic Fe$_3$O$_4$ and CFO reduces to

$$\Delta E = 16JS^2 \cos(\theta) + K(5 - \cos(2\theta)) - 16D_z S^2 \sin(\theta) \qquad (2a)$$

$$\Delta E = 16JS^2 \cos(\theta) + K(5 - \cos(2\theta)) - 16D_z S^2 \sin(\theta) \qquad (2b)$$

respectively [43].

In the above equations $J$, $K$, and $D_z$ are the isotropic exchange, single spin anisotropy, and DM exchange parameter, respectively. The procedure for deriving the above two expressions is given in Ref. 43.

Figure 7 & 11 plot the DFT penalty energy as a function of tetrahedral site canting for Fe$_3$O$_4$ and CFO, respectively. The fitting shows that DM term balances the energy cost associated with isotropic exchange and anisotropy term due to spin canting. The magnetic coupling parameters $J$, $K$ and $D_z$ for Fe$_3$O$_4$ and CFO are given in [Table 3, Ref. 43]. Thus, though DM can play a role in stabilizing a canted spin structure at the surface due to broken symmetry in the absence of applied magnetic field, the effect of Zeeman term can be visualized as further propagation of canted regions inside the nanoparticles which is balanced by the competing energy terms i.e. isotropic exchange and anisotropic energy and define the shell thickness as discussed earlier [37].

### B. CFO nanoparticles

In contrast to $Fe_3O_4$, the NCM calculation for CFO shows a smaller difference in cohesive energy or penalty energy between different canting configuration which is of the order of 0.001 meV compared to 1 meV in case of $Fe_3O_4$. This indicates that in CFO, the surface spins have the freedom to tilt along all possible directions [Fig. S5, Table 2, Ref. 43]. The various canting configuration and their relative energies are given in Table 2(a) [43]. The most stable configuration is 17° (both $T_d$ and $O_h$) and the corresponding moment per unit cell volume is 4.12 $\mu_B$. This particular configuration has two advantages, i.e. both the total energy and Zeeman energy savings due to maximum magnetic moment per formula unit under a magnetic field over other combination. This is in close agreement with the earlier PASANS method where a canting angle of 17° and 33° have been predicted with the help of the energy balance model at 300 and 77 K, respectively. Kindly note that in the absence of magnetic field the DM term will stabilize canting configurations of ($T_d$ 17° $O_h$ 17°) having the most stable energy value.

Theoretical EELS spectra along with DOS for the usual ferrimagnetic spin configurations are shown in Figure 8. The fine features corresponding to the Fe tetrahedral and octahedral atomic site contributions in the Fe-$L_3$ spectra are marked with different colors. As Co is only in the octahedral site, therefore the fine features in Co-$L_3$ spectra will have contributions only from Co $O_h$ DOS. Figure 9 shows how the distributions of DOS and discrete nature of peaks changes between the usual ferrimagnetic and the most stable (17°) spin canting configurations. For other canting configurations see Ref. 43. The theoretical changes are similar to $Fe_3O_4$ case and a quick distinction can be made only between the canted vs. ferrimagnetic configurations, but a comparison between various canting configurations is not a simple task to perform and can be the prospect of future work.

Experimental Fe-$L_3$ and Co-$L_3$ spectra from two different regions of nanoparticles and at two different temperatures are shown in Figure 10. The contributions from both tetrahedral and octahedral site contributions in case of Fe and octahedral site contributions in case of Co are indicated with arrows. CFO does not undergo any structural transitions at low temperature unlike $Fe_3O_4$ case; therefore, single structural model is sufficient to describe all the experimental spectra. Differences in peaks can only be observed between the core and edge regions of the CFO nanocrystal for Co-$L_3$ but not for Fe-$L_3$ spectra [Fig.10]. The features in Fe-$L_3$ suggests uniform canting of Fe spins throughout the nanocrystal but a core-shell morphology for Co spins. This is a very important finding and demonstrates the capability of HREELS technique in identifying atom-specific spin configurations, which may not be possible by neutron based techniques. The theoretical calculations also suggest the formation of such configuration due to the small energy difference compared to other spin canting geometry. From the experimental results we find essentially similar features in terms of fine structures between the core and surface regions of the nanoparticles. Therefore, the EELS based investigation is in agreement with earlier findings of randomly oriented spin canting structure in case of CFO based on experimental PASANS and an analysis based on an energy based model [38]. A value close to 17° is found to be the stable configuration at 300 K temperature.

The shell thickness is determined approximately by sliding the rectangular slice with various widths below 1 nm and observing for discernable changes in the spectra. The changes in the spectra are between the core-like and edge-like features. This gives us approximately a shell thickness of 2 (±0.2) and 1.2 ((±0.2) nm for $Fe_3O_4$ at 300 and 77 K, respectively, and 1.8 ((±0.2) nm for Co atom only shell thickness for CFO. Figure 12 provides a schematic of the spin canting geometry for the two different nanocrystals.

### IV.     Conclusions

In conclusion, we have developed a soft experimental technique based on HREELS to probe the magnetic structure in magnetic nanoparticles with high spatial and energy resolution. The technique enables recording HREELS spectra with high signal to noise ratio without causing damage to the specimen. The technique has been utilized to investigate surface spin canting in both $Fe_3O_4$ and CFO nanoparticles. The overall results are in good agreement with the previously reported polarized neutron based technique but the first principle calculations have helped us to narrow down the possible canting angles for $Fe_3O_4$ ($T_d$, $O_h$ tilts 40°, 40°) and CFO ($T_d$, $O_h$ tilts 17°, 17°). The role of DM interaction is also discussed and found to stabilize the spin canting structure at the surface and the Zeeman energy term aids in forming the canted shell thickness upon balancing with the competing energy terms. These results represent an extension of the HREELS technique to probe magnetic spin canting in low dimensional systems and can be further expanded to address various other problems at the nanometer and atomic plane resolution length scale.

**Acknowledgement**

The authors at ICMS, JNCASR sincerely thanks Prof. C. N. R. Rao for his constant support and advanced electron microscopy facility for this research. D. S. N. acknowledges CSIR India for a Ph.D fellowship. The work at the University of Alabama was supported by the National Science Foundation under Grant No. CHE-1508259.

**References**

[1] C. C. Berry and A. S. G. Curtis, J. Phys. D: Appl. Phys. **36**, R198 (2003).

[2] K. E. Scarberry, E. B. Dickerson, J. F. McDonald, and Z. J. Zhang, J. Am. Chem. Soc. **130**, 10258 (2008).

[3] J. H. Lee, J. T. Jang, J. S. Choi, S. H. Moon, S. H. Noh, J. W. Kim, J. G. K. Kim, S. Kim, K. I. Park, and J. Cheon, Nat. Nanotech. **6**, 418 (2011).

[4] J. Xie, K. Chen, and X. Chen, Nano Res. **2**, 261 (2009).

[5] D. Bader, Rev. Mod. Phys. **78**, 1 (2006).

[6] N. Bao and A. Gupta Encyclopedia of Inorganic and Bioinorganic Chemistry ed. R. A. Scott, John Wiley: DOI: 10.1002/9781119951438.

[7] T. Hyeon, Chem. Comm. **8**, 927 (2003).

[8] F. Li, D. Zhi, Y. Luo, J. Zhang, X. Nan, Y. Zhang, W. Zhou, B. Qiu, L. Wen, and G. Liang, Nanoscale **8**, 2826 (2016).

[9] Q. A. Pankhurst, J. Connolly, S. K. Jones, and J. Dobson, J. Phys. D: Appl. Phys. **36,** R167 (2003).

[10] S. Sun, H. Zeng, D. B. Robinson, S. Raoux, P. M. Rice, S. X. Wang, and L. Guanxiong, J. Am. Chem. Soc. **126**, 273 (2004).

[11] K. Ramasamy, D. Mazumdar, Z. Zhou, Y. Hsiang, A. Wang, and A. Gupta, J. Am. Chem. Soc. **133**, 20716 (2011).

[12] Y. P. Cai, K. Chesnel, M. Trevino, A. Westover, R. G. Harrison, J. M. Hancock, S. Turley, A. Scherz, A. Reid, B. Wu, C. Graves, T. Wang, T. Liu, and H. Dürr, J. Appl. Phys. **115**, 17B537 (2014).

[13] N. Bao, L. Shen, Y. Wang, P. Padhan, and A. Gupta, J. Am. Chem. Soc. **129**, 12374 (2007).

[14] S. Linderoth, P. V. Hendriksen, F. Bardker, S. Wells, K. Davies, S. W. Charles, and S. Moup, J. Appl. Phys. **75**, 6583 (1994).

[15] M. P. Morales, C. J. Sernaz, F. Bødkerx, and S Mørupx, J. Phys.: Condens. Matter **9**, 5461 (1997).

[16] F. T. Parker, M. W. Foster, D. T. Margulies, and A. E. Berkowitz, Phys. Rev. B **47**, 7885 (1993).

[17] S. Kubickova, D. Niznansky, M. P. Morales Herrero, G. Salas, and J. Vejpravova, Appl. Phys. Lett. **104**, 223105 (2014).

[18] J. M. D. Coey, Phys. Rev. Lett. **27**, 17 (1971).

[19] S. Laurent, D. Forge, M. Port, A. Roch, C. Robic, L. V. Elst, and R. N. Muller, Chem. Rev. **108**, 2064 (2008).

[20] A. Lu, E. L. Salabas, and F. Schuth, Angew. Chem. Int. Ed., **46**, 1222 (2007).

[21] H. Deng, X. Li, Q. Peng, X. Wang, J. Chen, and Y. Li, Angew. Chem. **117**, 2842 (2005).

[22] N. A. Frey, S. Peng, K. Cheng, and S. Sun, Chem. Soc. Rev. **38**, 2532 (2009).

[23] D. L. Huber, Small **1**, 482 (2005).

[24] A. G. Kolhatkar, A. C. Jamison, D. Litvinov, R. C. Willson, and T. R. Lee, Int. J. Mol. Sci. **14**, 15977 (2013).

[25] J. Park, K. An, Y. Hwang, J. G. Park, H. J. Noh, J. Y. Kim, J. H. Park, N. M. Hwang, and T. Hyeon, Nature Letter **3**, 891 (2004).

[26] D. S. Negi, B. Loukya, K. Ramasamy, A. Gupta, and R. Datta, Appl. Phys. Lett. **106**, 182402 (2015).

[27] J. Verbeeck, P. Schattschneider, S. Lazar, M. Stoger-Pollach, S. Loffler, A. Steiger-Thirsfeld, and G. Van Tendeloo, Appl. Phys. Lett. **99**, 203109 (2011).

[28] F. Gazeau1, F. Boue, E. Dubois, and R. Perzynski, J. Phys.: Condens. Matter **15,** S1305 (2003).

[29] P. Schattschneider, I. Ennen, S. Löffler, M. Stöger-Pollach, and J. Verbeeck, J. Appl. Phys. 09D311 **107**, (2010).

[30] B. Loukya, X. Zhang, A. Gupta, and R. Datta, J. Magn. Magn. Mater. **324**, 3754 (2012).

[31] B. Loukya, D. S. Negi, K. Dileep, N. Pachauri, A. Gupta, and R. Datta, Phys. Rev. B **91**, 134412 (2015).

[32] E. O. Wollan and W. C. Koehler, Phys. Rev. **100**, 545 (1995).

[33] J. Verbeeck, H. Tian, and P. Schattschneider, Nature Letter **467**, 301 (2010).

[34] A. Wiedenmann, Physica B **356**, 246 (2005).

[35] S. Disch, E. Wetterskog, R. P, Hermann, A. Wiedenmann, U. Vainio, G. Salazar-Alvarez, L. Bergstrom, and T. Bruckel, New Journal of Phys. **14**, 013025 (2012).

[36] V. E. Dmitrienko, E. N. Ovchinnikova, S. P. Collins, G. Nisbet, G. Beutier, Y. O. Kvashnin, V. V. Mazurenko, A. I. Lichtenstein, and M. I. Katsnelson, Nature Phys. **10**, 202 (2014).

[37] K. L. Krycka, J. A. Borchers, R. A. Booth, Y. Ijiri, K. Hasz, J. J. Rhyne, and S. A. Majetich, Phys. Rev. Lett. **113**, 147203 (2014).

[38] K. Hasz, Y. Ijiri, K. L. Krycka, J. A. Borchers, R. A. Booth, S. Oberdick, and S. A. Majetich, Phys. Rev. B **90**, 180405(R) (2014).

[39] J. Salafranca, J. Gazquez, N. Pérez, A. Labarta, S. T. Pantelides, S. J. Pennycook, X. Batlle, and M. Varela, Nano Lett. **12**, 2499 (2012).

[40] R. Laskowski, G. K. H. Madsen, P. Blaha, and K. Schwarz, Phys. Rev. B **69**, 140408 (R) (2004).

[41] A. Garcia, A. M. Raya, M. M. Mariscal, R. Esparza, M. Herrera, S. Molina G.Scavello, P. Galindo, M. J. Yacaman, and A. Ponce, Ultramicroscopy **146**, 33 (2014).

[42] N. Bao, L. Shen, W. An, P. Padhan, C. H. Turner, and A. Gupta, Chem. Mater. **21**, 2345 (2009).

[43] Supplementary material for the detail of particle damage by monoprobe, STEM probe, novel HR-EELS technique, various canting configuration considered for first principle calculation, DOS, and derivation of spin Hamiltonian for penalty energy with fitted exchange and anisotropy parameter.

[44] P. Blaha, K. Schwarz, G. K. H. Madsen, D. Kvasnicka, and J. Luitz, WIEN2k: An Augmented Plane Wave + Local Orbitals Program for Calculating Crystal Properties (Karlheinz Schwarz, Techn. Universität Wien, Austria, 2001).

[45] H. T. Jeng, G. Y. Guo, and D. J. Huang, Phys. Rev. Lett. **93**, 156403 (2004).

[46] N. M. Caffrey, D. Fritsch, T. Archer, S. Sanvito, and C. Ederer, Phys. Rev. B **87**, 024419 (2013).

[47] H. T. Jeng and G. Y. Guo, Phys. Rev. B **65**, 094429 (2002).

[48] K. Dileep, B. Loukya, N. Pachauri, A. Gupta, and R. Datta, J. Appl. Phys. **116**, 103505 (2014).

[49] R. F. Egerton, Rep. Prog. Phys. **72**, 016502 (2009).

[50] I. Dzyalonshinsky, J. Phys. Chem. Solids **4**, 241 (1958).

[51] T. Morial, Phys. Rev. Lett. **120**, 91 (1960).

[52] Y. S. Hou, H. J. Xiang, and X. G. Gong, New Journal of Phys. **18**, 043007 (2016).

[53] P. Liu, S. Khmelevskyi, B. Kim, M. Marsman, D. Li, X. Q. Chen, D. D. Sarma, G. Kresse, and C. Franchini, Phys. Rev. B **92**, 54428 (2015).

[54] M. Bode, M. Heide, K. V. Bergmann, P. Ferriani, S. Heinze, G. Bihlmayer, A. Kubetzka, O. Pietzsch, S. Blugel, and R. Wiesendanger, Nature Letter **447**, 190 (2007).

[55] C. Weingart, N. Spaldin, and E. Bousquet, Phys. Rev. B **86**, 094413 (2012).

[56] P. W. Ma and S. L. Dudarev, Phys. Rev. B **91**, 054420 (2015).

[57] H. J. Xiang, E. J. Kan, Su-Huai Wei, M. H. Whangbo, and X. G. Gong, Phys. Rev. B **84**, 224429 (2011).

[58] B. H. Kim and B. I. Min, New Journal of Phys. **13**, 073034 (2011).

**Figures:**

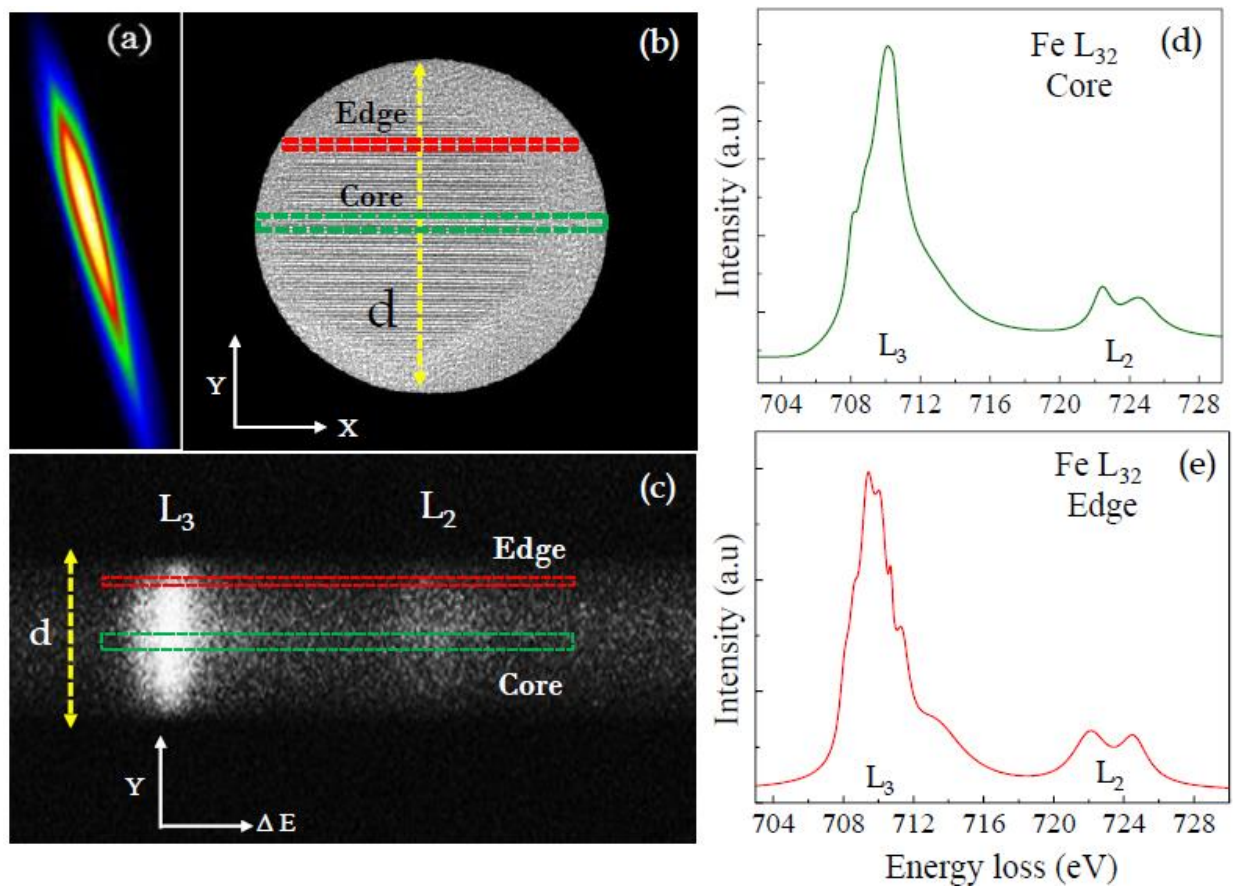

**FIG. 1.** Details of the novel soft HR-EELS based technique with one atomic plane resolution. (a) Image of mono probe that is used to illuminate the nanoparticle sample, (b) example atomic resolution image of ~ 10 nm $Fe_3O_4$ nanoparticle as seen through 1 mm GIF entrance aperture in the image mode of spectrometer. In the spectroscopy mode the entire image of (b) is dispersed as Y vs ΔE as shown in (c) every Y data points in figure (c) has all the corresponding X data points projected in it. Rectangular slice tool is used to extract spatially resolved HR-EELS spectra from the (d) core and (e) edge of the nanoparticle with high signal to noise ratio.

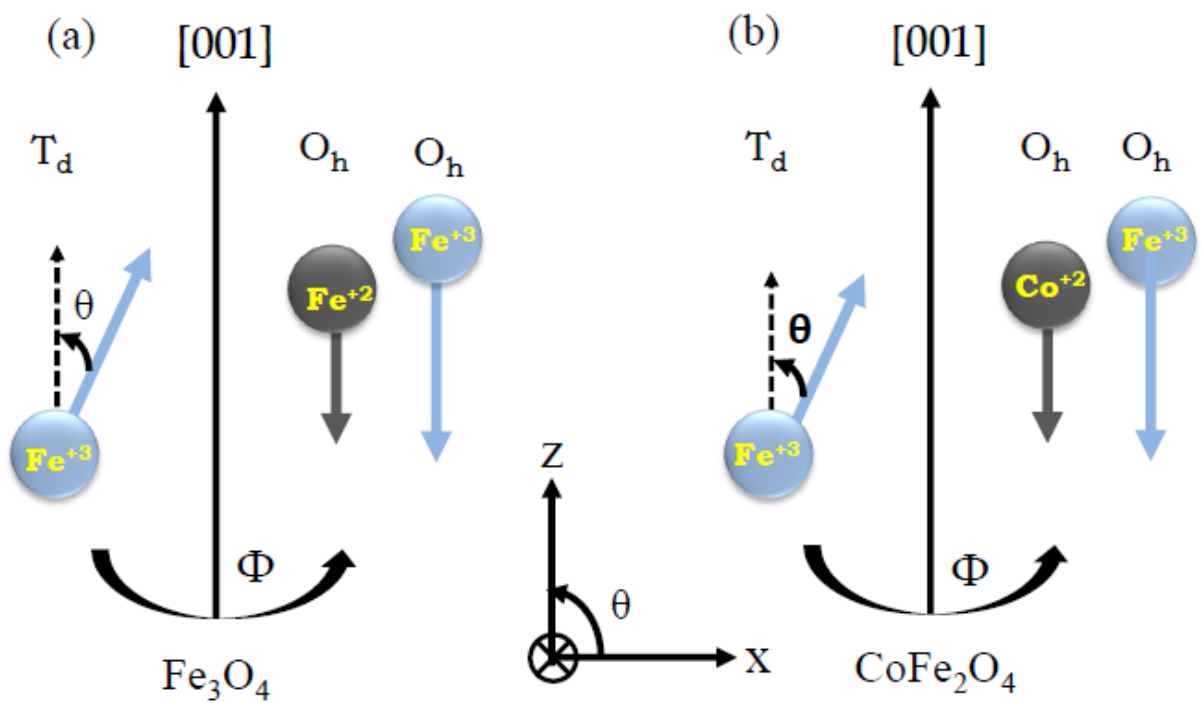

**FIG. 2.** Schematic representation of spin canting geometry for (a) Fe$_3$O$_4$ and (b) CFO nanocrystals with respect to the corresponding easy axis. θ is the spin canting angle with respect to easy axis and Φ is the azimuthal canting angle about easy axis. For example, only T$_d$ canting is shown however, any combination of canting in terms of θ, Φ is possible between the three different magnetic ions in the inverse spinel structure.

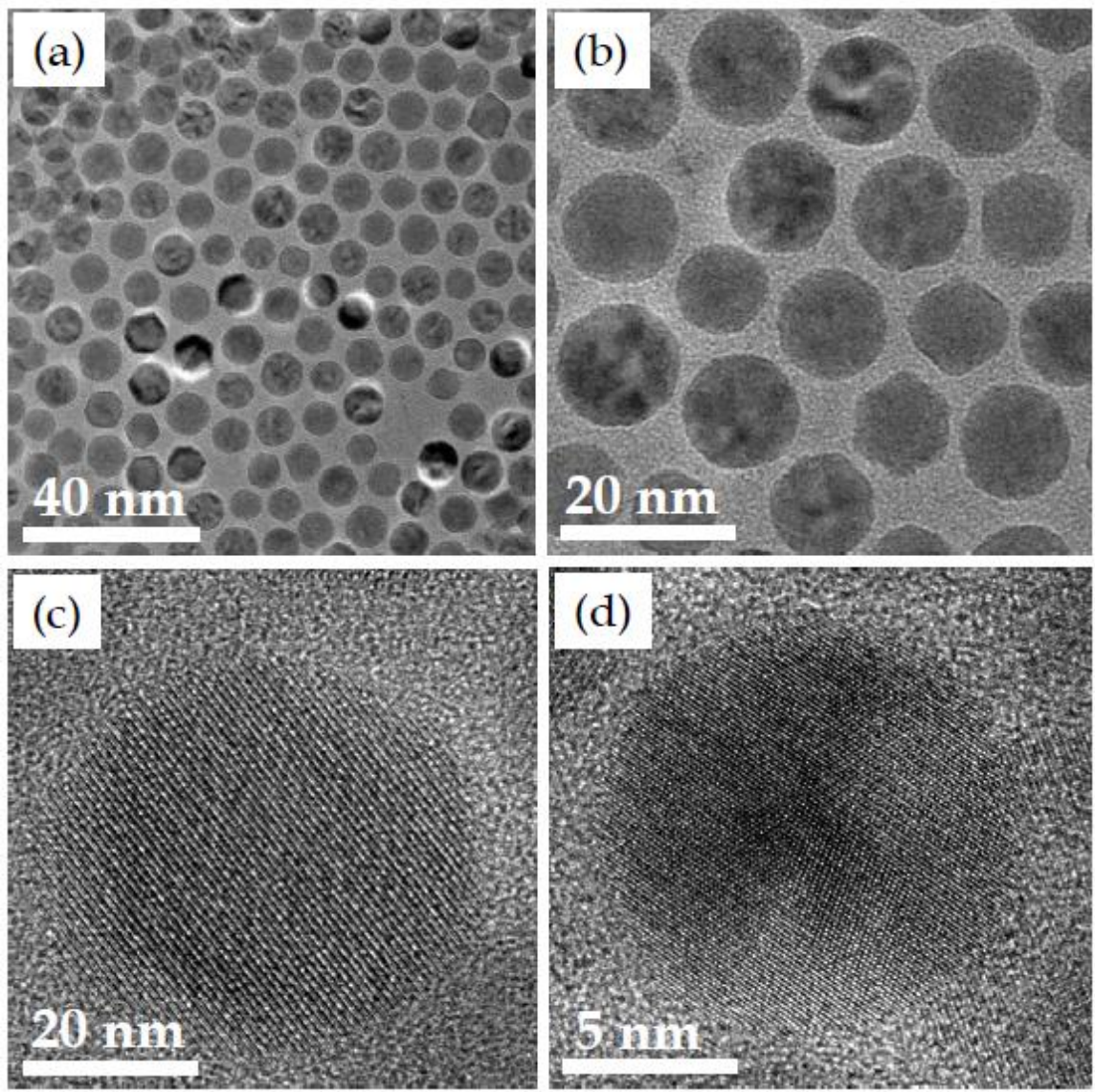

**FIG. 3.** (a) & (b) low magnification and (c) & (d) high resolution TEM images of Fe$_3$O$_4$ and CFO nanoparticles, respectively. HRTEM images show that the nanoparticles are single crystalline and free from any visible crystallographic defects.

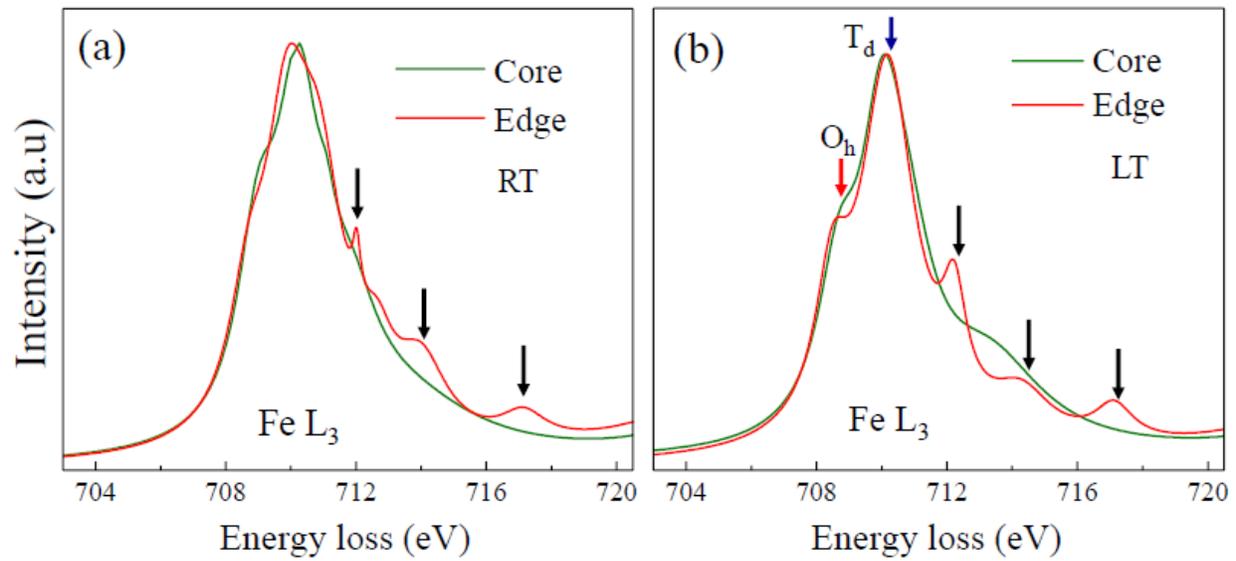

**FIG. 4.** Experimental $L_3$ spectra of $Fe_3O_4$ recorded (a) at room temperature and (b) at liquid nitrogen temperature (77 K). The spectra from the core and edge regions of the nanoparticles are colored with green and red, respectively. Dominating feature from $T_d$ and $O_h$ atomic site are marked. Fine features become more discrete in shape in the spectra taken from the edge region of nanocrystals.

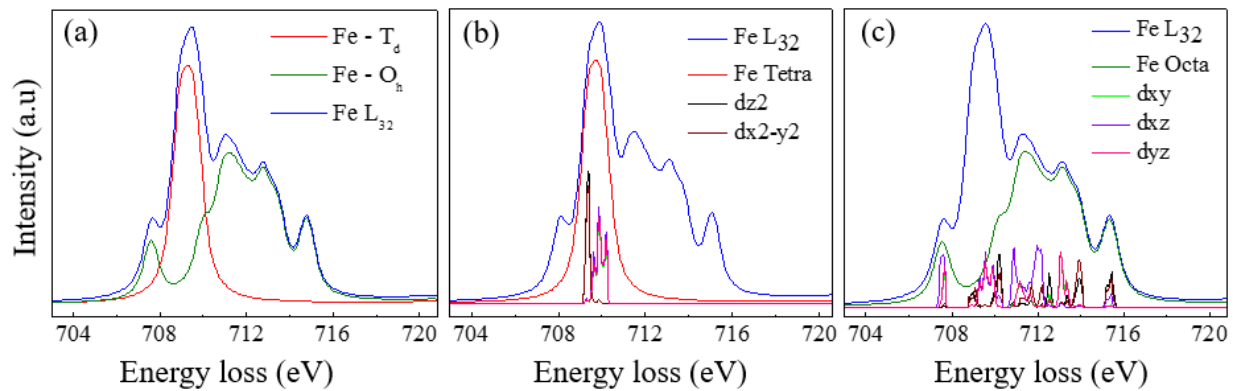

**FIG. 5.** Simulated Fe $L_{3,2}$ spectra in $Fe_3O_4$ shown with blue color. (a) $T_d$ and $O_h$ atomic contribution, (b) $T_d$ contribution along with partial density of states, and (c) $O_h$ contribution along with partial density of states to the averaged spectra are depicted with the aid of various color plots.

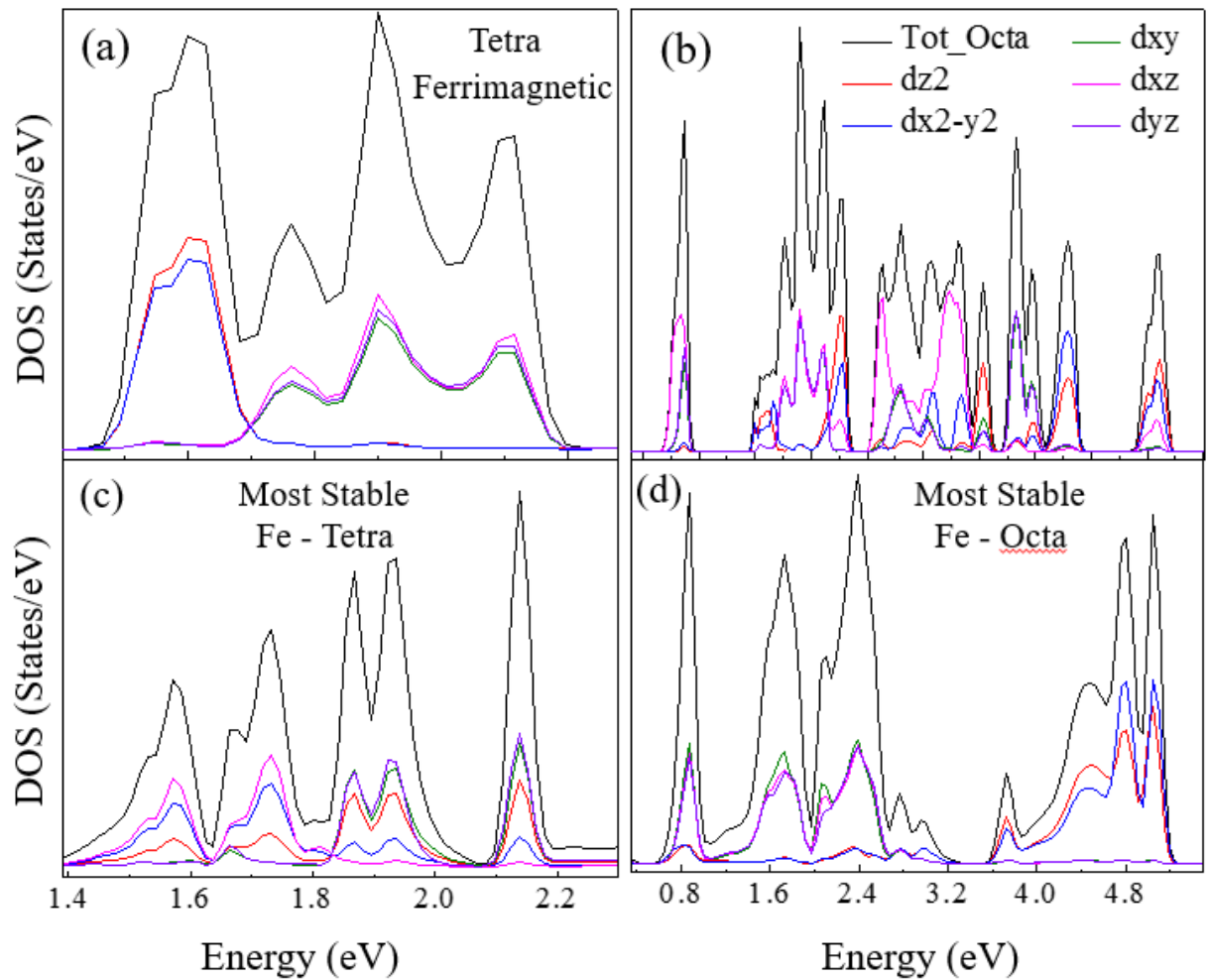

**FIG. 6.** Total density of unoccupied states (black line) with partial contributions for (a) $T_d$ and (b) $O_h$ Fe atoms in $Fe_3O_4$ with ferrimagnetic spin arrangement as calculated in WIEN NCM code. The corresponding total density of state and partial contribution for most stable configuration for $T_d$ and $O_h$ atoms are given in (c) and (d), respectively.

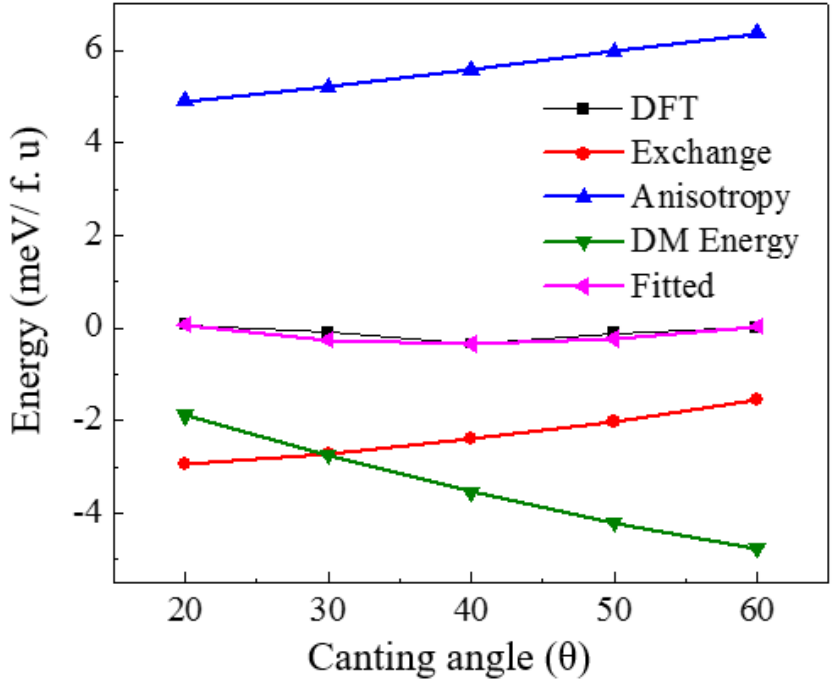

**FIG. 7.** Plot of the spin dependent penalty energy cost calculated by DFT as a function of $T_d$ canting angle for $Fe_3O_4$. The penalty energy cost is split into three competing energy terms: isotropic exchange energy (red line), anisotropy energy (blue line), and DM interaction energy (green line).

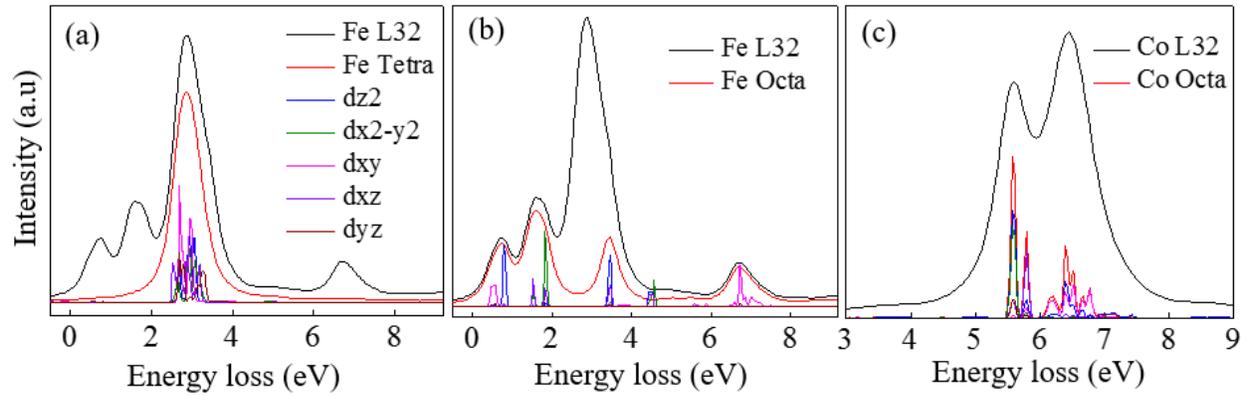

**FIG. 8.** Simulated Fe L$_3$, spectra in CFO shown with blue color. (a) T$_d$ contribution along with partial density of states, and (b) O$_h$ contribution along with partial density of state to the averaged spectra are depicted with the help of various color plots. (c) Co L$_3$ spectra with partial contributions of Co O$_h$ density of states.

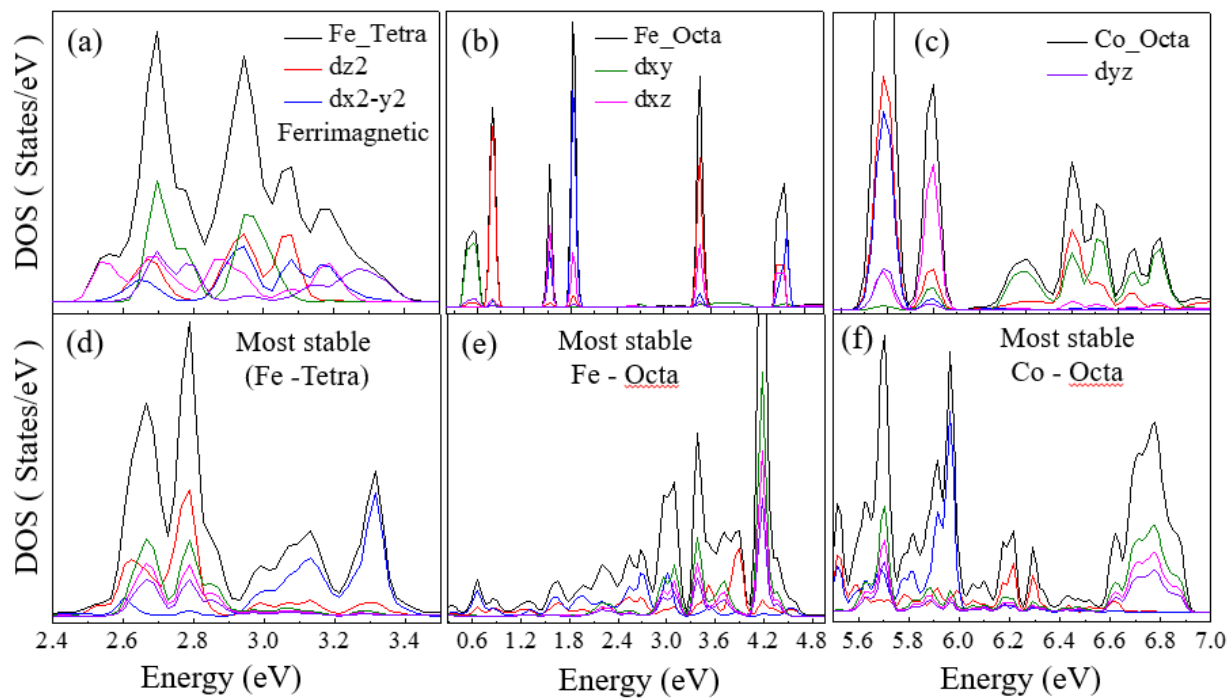

**FIG. 9.** Total density of unoccupied states (black line) with partial contributions for (a) $O_h$ Co, (b) $O_h$ Fe, and (c) $T_d$ Fe in CFO with ferrimagnetic spin arrangement as calculated in WIEN NCM code. The corresponding total density of state and partial contribution for most stable configuration corresponding to (a), (b), and (c) are given in (d), (e), and (f), respectively.

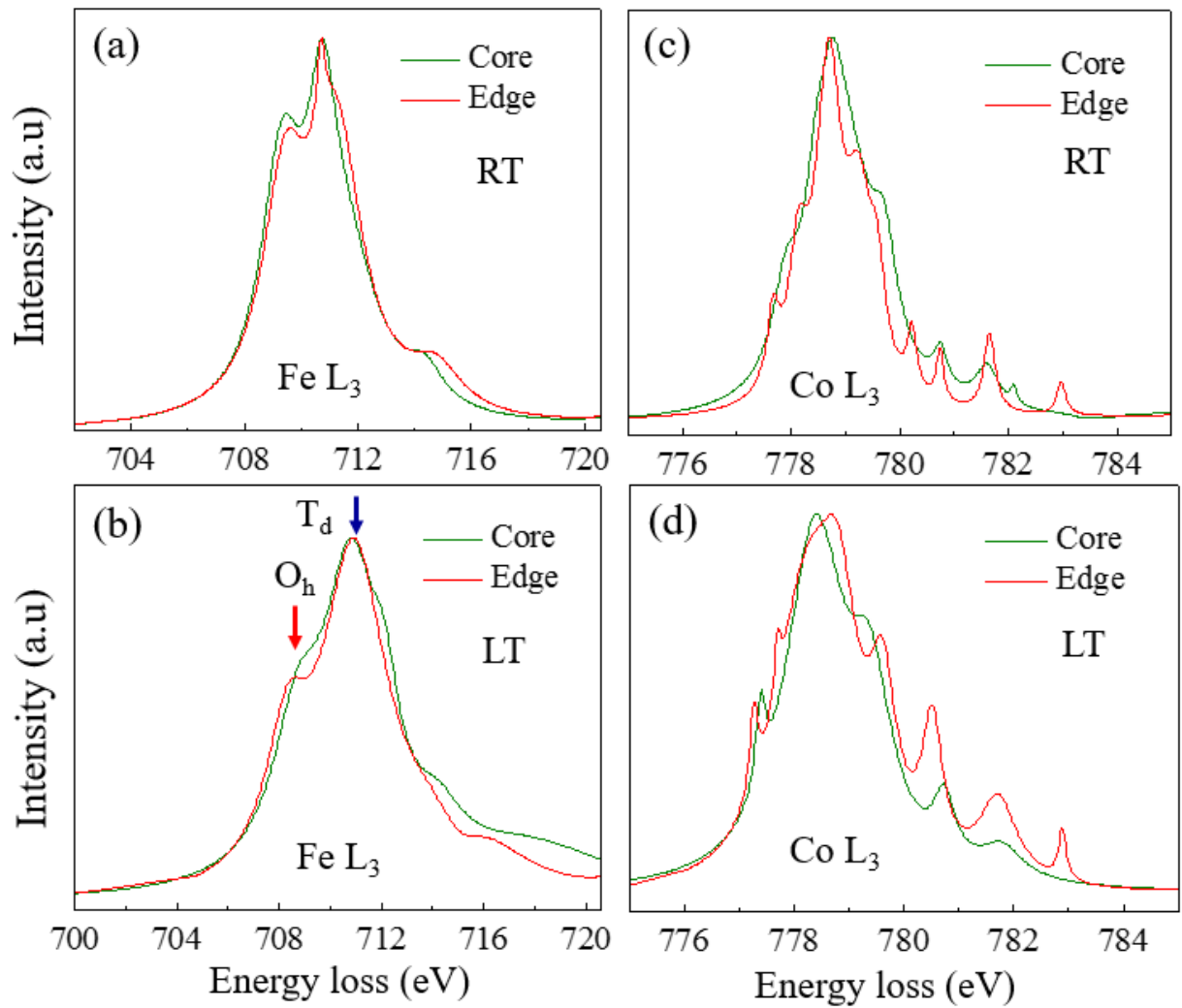

**FIG. 10.** Experimental L$_3$ spectra of CFO recorded (a) & (c) at room temperature and (b) & (d) at liquid nitrogen temperature (77 K) for Fe and Co atoms, respectively. The spectra from core and edge of nanoparticles are colored with green and red, respectively. Dominating feature from T$_d$ and O$_h$ atomic site are marked. Kindly note the fine features are only shaper for Fe atoms but not for Co atoms, suggesting possible formation of uniformly oriented spin canting configuration for Fe atoms but core shell morphology for Co atoms.

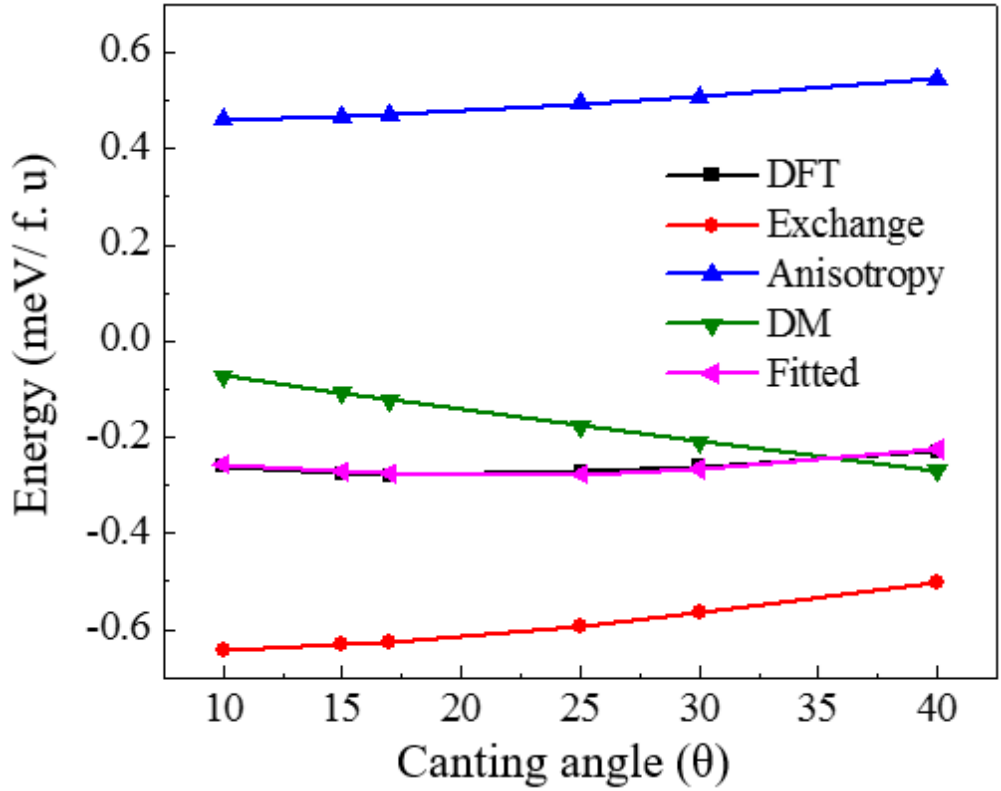

**FIG. 11.** Plot of the spin dependent penalty energy cost calculated by DFT as a function of $T_d$ canting angle for CFO. The penalty energy cost is split into three competing energy terms: isotropic exchange energy (red line), anisotropy energy (blue line), and DM interaction energy (green line).

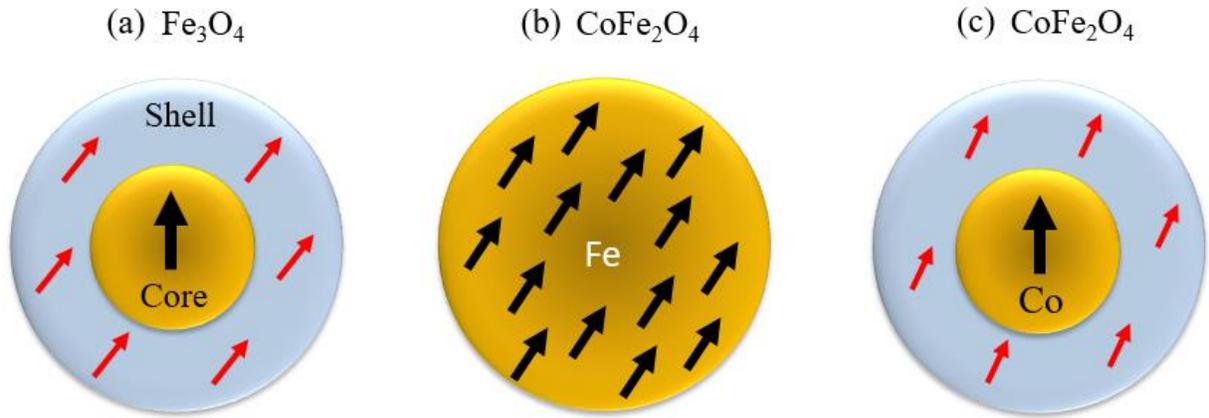

**FIG. 12.** Schematic of general spin canting geometry for (a) $Fe_3O_4$ and (b)&(c) for CFO nanoparticles. In case of CFO, Co atom forms the core shell canting configuration but Fe forms the uniformly oriented canting configuration.

# Surface spin canting in $Fe_3O_4$ and $CoFe_2O_4$ nanoparticles probed by high resolution electron energy loss spectroscopy

D. S. Negi,[1,2] H. Sharona,[1,2] U. Bhat,[1,2] S. Palchoudhury,[4] A. Gupta[3] and R. Datta[1,2]*

[1]*International Centre for Materials Science, Jawaharlal Nehru Centre for Advanced Scientific Research, Bangalore 560064, India.*

[2]*Chemistry and Physics of Materials Unit, Jawaharlal Nehru Centre for Advanced Scientific Research, Bangalore 560064, India.*

[3]*Center for Materials for Information Technology, University of Alabama, Tuscaloosa, Alabama 35487, USA.*

[4]*Civil & Chemical Engineering, University of Tennessee at Chattanooga, Chattanooga, Tennessee 37403, USA.*

*Corresponding author E-mail: ranjan@jncasr.ac.in

## 1. A new soft technique based on EELS

We have developed a soft technique based on EELS in order to overcome beam damage and at the same time achieving high spatial resolution with high signal to noise ratio. Among known techniques available in TEM, the obvious choice is STEM-EELS combination to achieve high spatial resolution at atomic and subatomic length scale. However, due to high probe current of a typical STEM probe it is very difficult to acquire high quality core loss spectra which generally appears at high energy loss and requires sufficient exposure time to obtain a clear spectrum over noise before the sample is damaged. For most of the samples, damage is caused by drilling holes from beam exposure. Figure S1(a) shows STEM probe with spot size 9 (without x-FEG gun) drill holes in $CuCr_2S_4$ nanoparticles within few seconds of exposure in a FEI TITAN microscope. For this sample even a nanoprobe is found to drill holes quickly [1]. Moreover, it is also not convenient to work in a STEM mode in combination with a gun monochromator system to achieve high energy resolution other than spatial resolution. Therefore, we have made use of GIF entrance aperture as selected area aperture, objective magnification and nanoprobe magnification combination to achieve high spatial resolution with high signal to noise ratio. We have two variants of this soft technique, first one for selecting areas within few square nanometers area and the second one to achieve single atomic plane resolution, depending on the need one can select. Generally, objective magnification and GIF aperture together can be used to select a very small area within few square nanometers, any loss in intensity due to objective magnification can be compensated by mono demagnification on the sample. Mono probe demagnification is an independent control and can be used to improve signal to noise ratio. This forms the first method of the technique (Figure S2). The added advantage of this technique is that electron beam will remain as parallel illumination, which will be extremely useful if one wants to exploit this technique for EMCD. In the second variant, the whole nanoparticle is projected through GIF with atomic resolution and from the $Y$ vs. $\Delta E$ (real space), equivalent to $q$ vs. $\Delta E$ (diffraction space) plot one can extract information at spatial resolution of one atomic plane with the help of slices as shown in Figure S3. This method allows collecting HREELS spectra with both high spatial and energy resolution without beam damage.

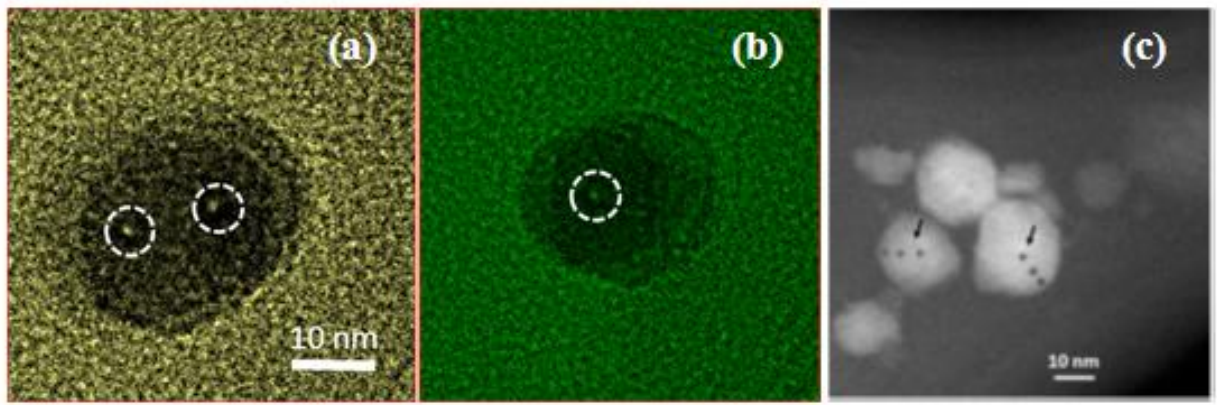

**Figure S1.** Focused electron beam drills holes in $CuCr_2S_4$ nanoparticles: (a) & (b) in nanoprobe mode and (c) in STEM mode.

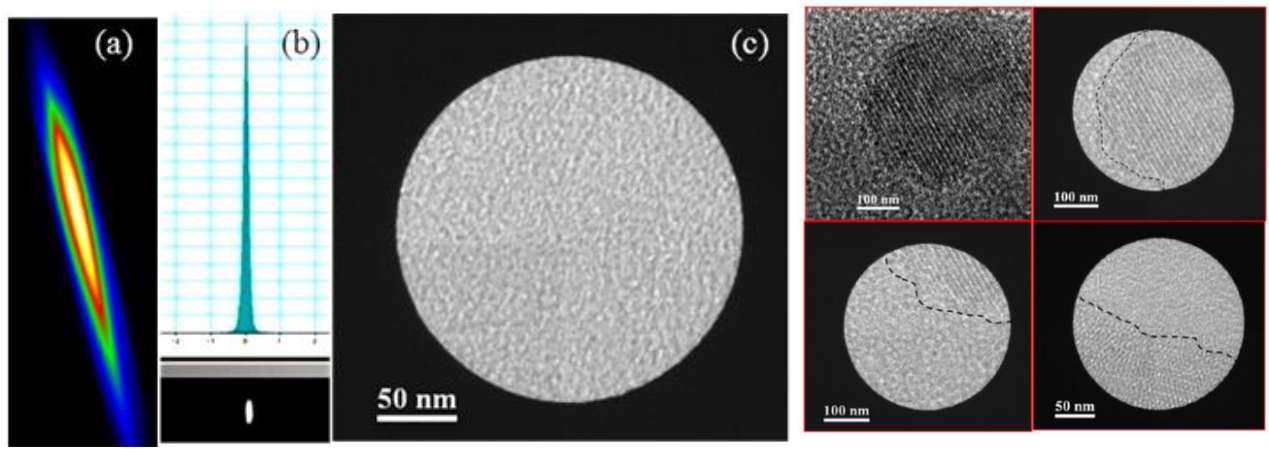

**Figure S2.** First variant of the soft technique where combination of nanoprobe demagnification, objective magnification and GIF aperture combination is used to acquire spectra with high spatial and energy resolution and at the same time avoiding beam damage. (a) Monoprobe, (b) energy resolution is measured from FWHM of the zero loss peak, (c) GIF entrance aperture and (d) example various regions of nanoparticles projected through GIF aperture.

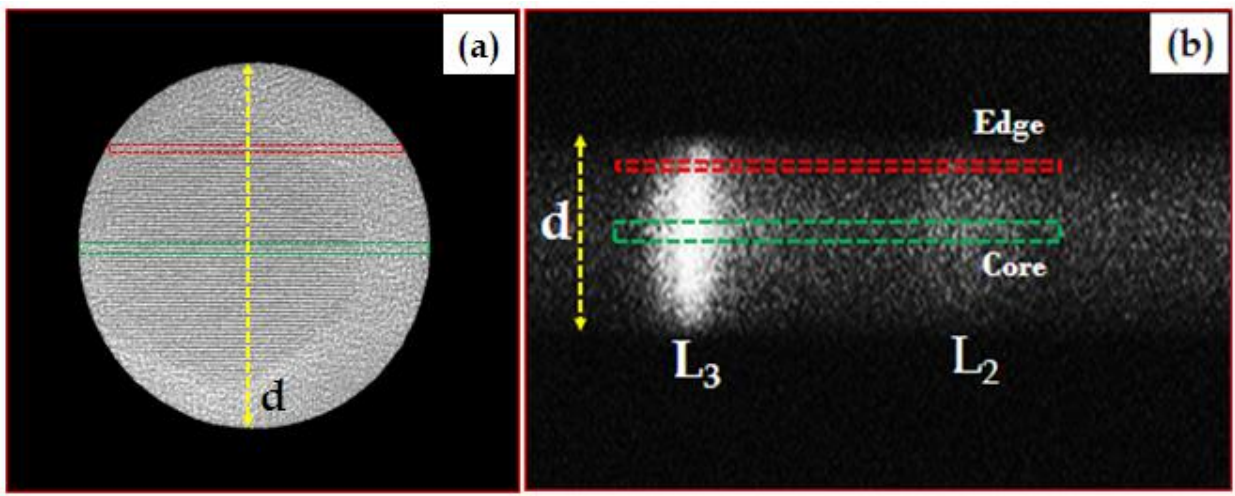

**Figure S3.** The second variant of the soft technique showing (a) projection of the entire nanoparticle with atomic plane resolution and (b) $Y$ vs. $\Delta E$ plot from which after calibration one can extract HREELS spectra at atomic plane resolution. See main text for the example spectra extracted from this dispersion using slice tool.

## 2. Details on the various spin canting geometry considered for theoretical calculation for $Fe_3O_4$ and $CoFe_2O_4$

All the regular ferrimagnetic calculations of $Fe_3O_4$ and $CoFe_2O_4$ (CFO) have been carried out using WIEN2k code [2]. For the calculation of various spin canting geometry, non collinear magnetism calculation is carried out using WIEN-NCM code as implemented within the same package. A magnetic supercell is created for WIEN-NCM calculation with the desired spin canting orientation at particular atomic sites. Example magnetic supercells with spin canting geometries are shown in Figure S4. Magnetic constraint calculation is carried out with U correction. The onsite potential U value is taken as 5 and 4.5 eV for $Fe_3O_4$ and CFO, respectively [3,4]. In order to evaluate the contribution of anisotropy and Dzyaloshinskii-Moriya (DM) interaction energy, which are essentially spin orbit interaction driven effect. Therefore, GGA+SOC+U was considered within the constraint spin first principal calculation. The various spin canting geometries and corresponding energy values for two different systems are summarized in Table I & II.

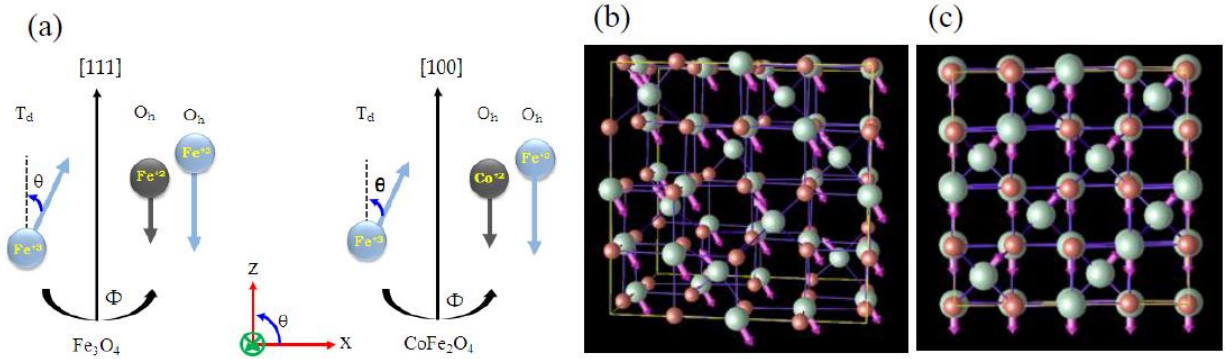

**Figure S4.** Spin canting geometry for non collinear magnetic calculation in WIEN NCM. (a) schematic canting geometry for $Fe_3O_4$ and CFO showing $T_d$ only tilt, where θ is canting angle with respect to the easy axis and Φ is azimuthal rotation about the easy axis. (b) schematic easy axis spin orientation along [111] directions, and (c) is the most stable ($T_d$, $O_h$, 40°, 40°) canting geometry for $Fe_3O_4$.

The cohesive energy plot for various spin canting geometry for Fe$_3$O$_4$ and CFO is shown in figure S5. Kindly note that the same can be plotted using total energy values as well, and for comparing relative energy it does not matter which representation is used as the difference in energy between various configurations will be same.

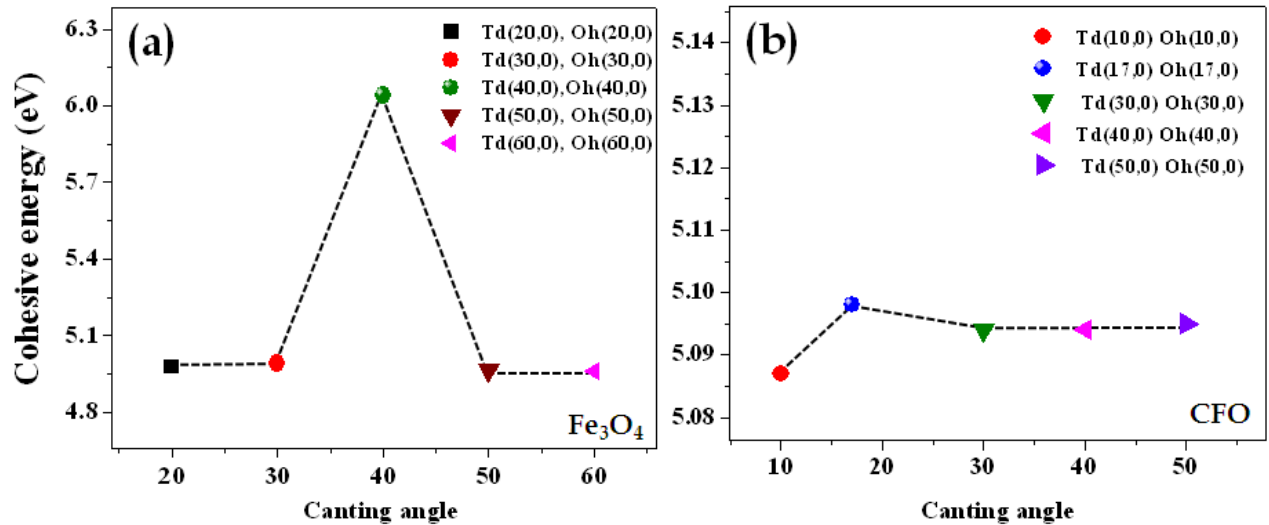

**Figure S5.** Cohesive energy plot representing relative stability of various spin canting geometry for (a) Fe$_3$O$_4$ and (b) for CFO. For various other spin canting geometry and their cohesive energy see Table 1 & 2.

Comparison of total DOS along with partial contributions between ferri magnetic, least stable canting configurations for both octa and tetra atoms in Fe$_3$O$_4$.

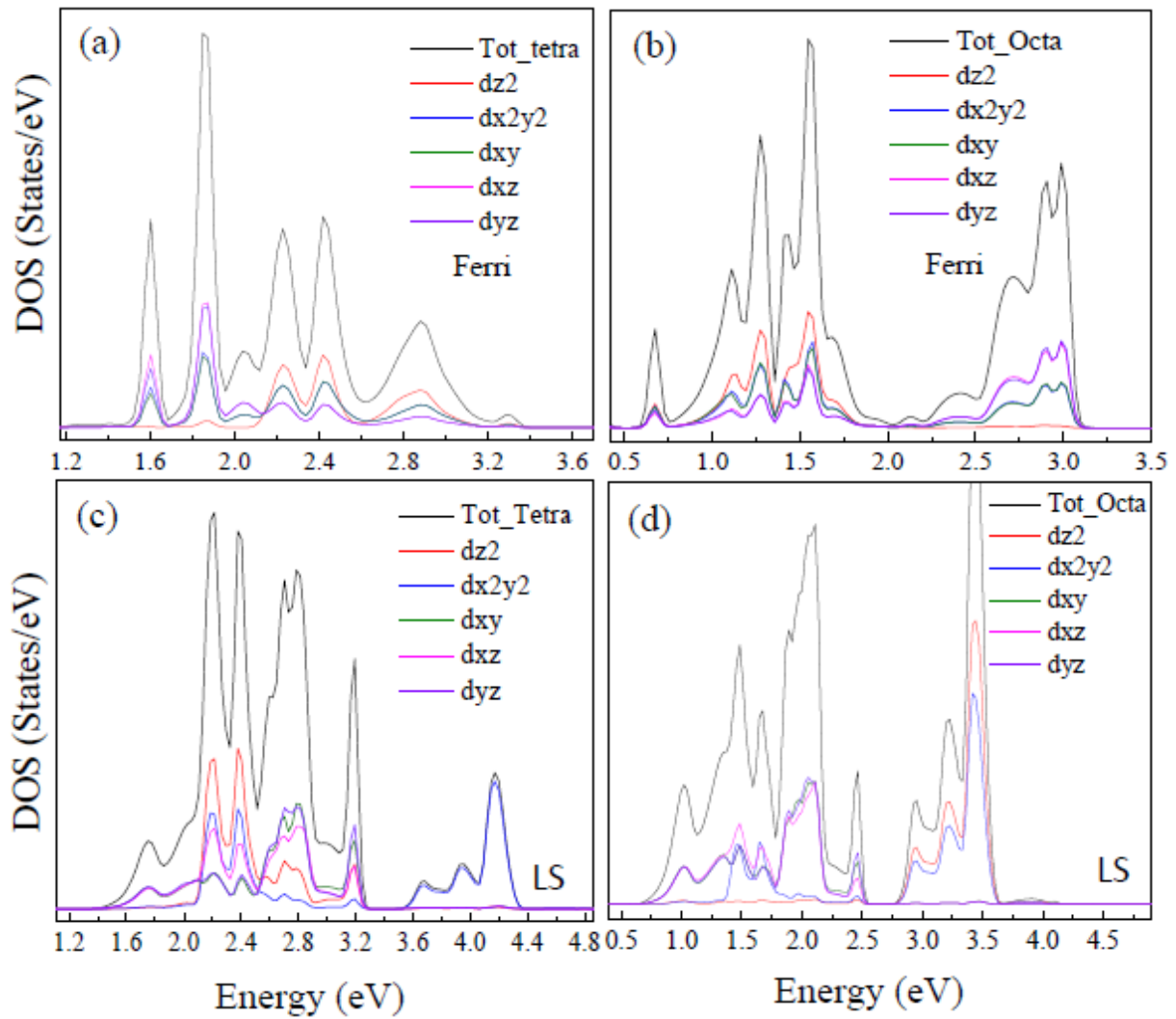

**Figure S6.** Total (a) octahedral and (c) tetrahedral DOS along with partial contributions for Fe$_3$O$_4$ ferrimagnetic configurations. (b) and (d) are the corresponding total and partial DOS for least stable spin configurations, respectively.

**Table 1.** (a) Various canting configuration, cohesive (per atom) and magnetic moment/ cell for $Fe_3O_4$. The ferrimagnetic and most stable configurations are highlighted. The ferrimagnetic and most stable configurations are highlighted. (b) Various azimuthal configurations for the most stable $T_d$ canting 40º.

**Table 1(a)**

| Canting Angle | $T_d$ Θ=0 Φ=0 $O_h$ Θ=0 Φ=0 [111] | $T_d$ Θ=20 Φ=0 $O_h$ Θ=0 Φ=0 | $T_d$ Θ=30 Φ=0 $O_h$ Θ=0 Φ=0 | $T_d$ Θ=40 Φ=0 $O_h$ Θ=0 Φ=0 | $T_d$ Θ=50 Φ=0 $O_h$ Θ=0 Φ=0 | $T_d$ Θ=60 Φ=0 $O_h$ Θ=0 Φ=0 |
|---|---|---|---|---|---|---|
| Spin Configuration | ↑↓ | ↗↓ | ↗↓ | ↗↓ | ↗↓ | ↗↓ |
| Cohesive Energy (eV)/atom | 6.04 | 4.97 | 4.99 | 6.02 | 4.96 | 4.98 |
| Moment $\mu_B$/cell | 4.59 | 4.35 | 4.97 | 5.24 | 4.96 | 4.95 |

**Table 1(b)**

| Canting Angle | $T_d$ $\Theta=40$ $\Phi=0$ $O_h$ $\Theta=0$ $\Phi=0$ | $T_d$ $\Theta=0$ $\Phi=0$ $O_h$ $\Theta=40$ $\Phi=0$ | $T_d$ $\Theta=0$ $\Phi=0$ $O_h$ $\Theta=40$ $\Phi=0$ | $T_d$ $\Theta=0$ $\Phi=0$ $O_h$ $\Theta=40$ $\Phi=0$ | $T_d$ $\Theta=40$ $\Phi=0$ $O_h$ $\Theta=40$ $\Phi=0$ | $T_d$ $\Theta=40$ $\Phi=90$ $O_h$ $\Theta=0$ $\Phi=0$ | $T_d$ $\Theta=40$ $\Phi=90$ $O_h$ $\Theta=40$ $\Phi=90$ | $T_d$ $\Theta=40$ $\Phi=90$ $O_h$ $\Theta=40$ $\Phi=0$ | $T_d$ $\Theta=40$ $\Phi=45$ $O_h$ $\Theta=40$ $\Phi=0$ | $T_d$ $\Theta=40$ $\Phi=45$ $O_h$ $\Theta=40$ $\Phi=0$ | $T_d$ $\Theta=40$ $\Phi=45$ $O_h$ $\Theta=40$ $\Phi=45$ |
|---|---|---|---|---|---|---|---|---|---|---|---|
| Spin Configuration | ↗↓ | ↑↓ | ↑↓ | ↑↓ | ↗↓ | ↗↓ | ↗↓ | ↗↓ | ↗↓ | ↗↓ | ↗↓ |
| Cohesive Energy (eV)/atom | 6.02 | 6.02 | 6.02 | 6.00 | 6.04 | 6.02 | 6.03 | 5.99 | 6.01 | 6.02 | 6.03 |
| Moment $\mu_B$/Cell | 4.38 | 5.07 | 5.07 | 4.99 | 4.47 | 4.34 | 4.44 | 4.48 | 4.37 | 4.55 | 4.54 |

**Table 2.** Various spin canting arrangement and associated moment and energy values for CFO. The ferrimagnetic and most stable configurations are highlighted.

**Table 2(a)**

| Canting Angle | $T_d$ $\Theta=0$ $\Phi=0$ $O_h$ $\Theta=0$ $\Phi=0$ [100] | $T_d$ $\Theta=10$ $\Phi=0$ $O_h$ $\Theta=10$ $\Phi=0$ | $T_d$ $\Theta=17$ $\Phi=0$ $O_h$ $\Theta=17$ $\Phi=0$ | $T_d$ $\Theta=17$ $\Phi=90$ $O_h$ $\Theta=17$ $\Phi=90$ | $T_d$ $\Theta=30$ $\Phi=0$ $O_h$ $\Theta=0$ $\Phi=0$ | $T_d$ $\Theta=30$ $\Phi=0$ $O_h$ $\Theta=30$ $\Phi=0$ | $T_d$ $\Theta=0$ $\Phi=0$ $O_h$ $\Theta=40$ $\Phi=0$ | $T_d$ $\Theta=40$ $\Phi=0$ $O_h$ $\Theta=40$ $\Phi=0$ | $T_d$ $\Theta=40$ $\Phi=0$ $O_h$ $\Theta=40$ $\Phi=90$ | $T_d$ $\Theta=50$ $\Phi=0$ $O_h$ $\Theta=50$ $\Phi=0$ |
|---|---|---|---|---|---|---|---|---|---|---|
| Spin Config. | ↑↓↓ | ↑↓↓ | ↑↓↓ | ↓↓↻↻ | ↑↓↓ | ↑↓↓ | ↑↓↓ | ↑↓↓ | ↑↓↻ | ↑↓↓ |
| Moment µB/cell | 4.08 | 4.38 | 4.12 | 4.02 | 3.99 | 4.06 | 3.92 | 4.08 | 4.02 | 3.95 |
| Cohesive Energy (eV)/atom | 5.10 | 5.087 | 5.098 | 5.098 | 5.094 | 5.092 | 5.094 | 5.094 | 5.096 | 5.095 |

**Table 2(b)**

| Canting Angle | $T_d$ $\Theta=30$ $\Phi=0$ $O_h$ $\Theta=0$ $\Phi=0$ | $T_d$ $\Theta=0$ $\Phi=0$ $O_h$ $\Theta=30$ (Cobalt) $\Phi=0$ | $T_d$ $\Theta=30$ $\Phi=0$ $O_h$ $\Theta=30$ $\Phi=0$ |
|---|---|---|---|
| Canting Configuration | ↑↓↓ | ↑↓↓ | ↑↓↓ |
| Moment $\mu_B$/cell | 4.60 | 4.64 | 4.06 |
| Cohesive Energy (eV)/atom | 5.033 | 5.038 | 5.093 |

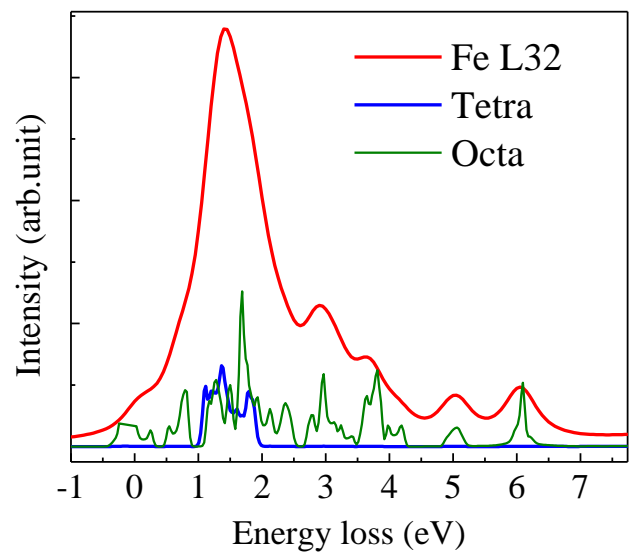

**Figure S7.** Simulated ELNES spectra of $Fe_3O_4$ monoclinic structure with Tetra and Octa DOS.

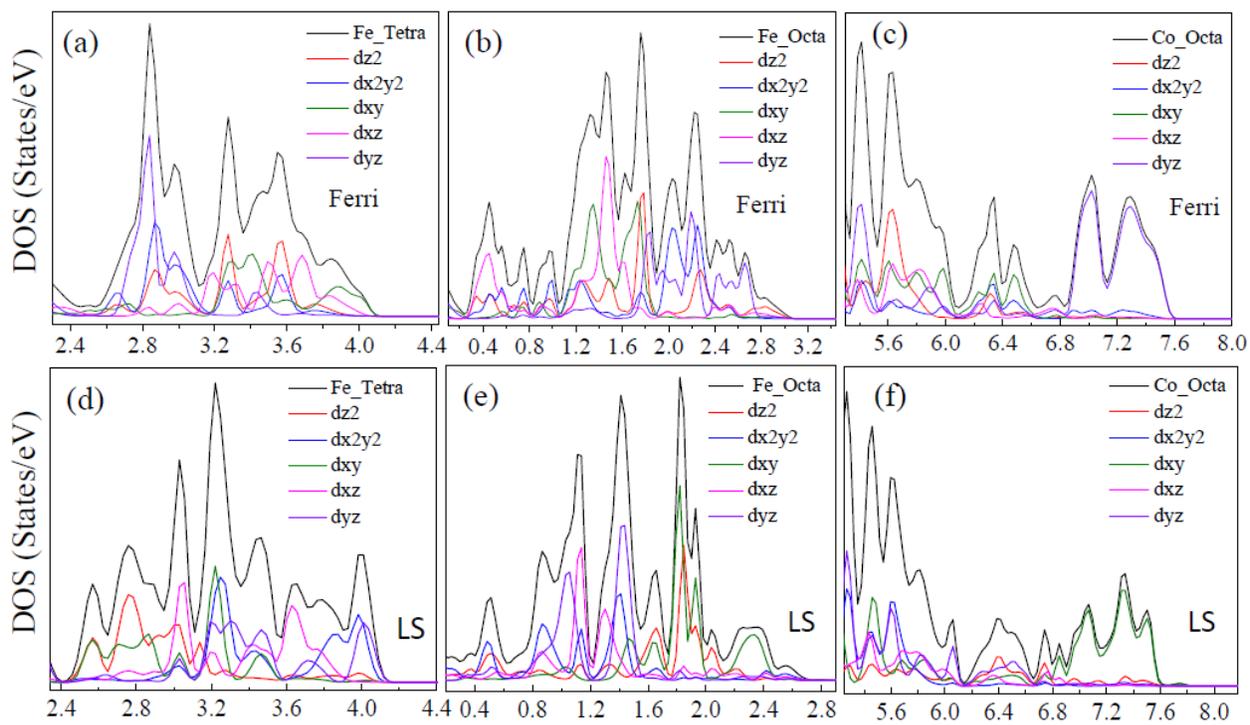

Figure S8 (a), (b), (c) are tetrahedral and octahedral partial DSO for ferrimagnetic CFO. (d), (e), (f) are the tetrahedral octahedral partial DOS for least stable spin canting configuration of CFO.

## 3. Derivation of spin Hamiltonian for $Fe_3O_4$ and CFO for fitting DM interaction term from the DFT penalty energy due to spin canting

In this section the expressions corresponding to various competing energy terms are given which have been used to fit the DFT penalty energy term under constraint magnetic calculation. Anisotropy and DM contributions are considered as the anisotropic exchange interactions which are due to the symmetry constraint spin orbit interaction. The additional energy cost is fitted with isotropic exchange, anisotropic energy, DM interaction or in other words can be mapped to the classical spin Hamiltonian as a penalty energy cost.

The penalty energy term $\Delta E_{spin}$ can be written as the sum of exchange interaction ($\Delta E_{ex}$), anisotropic energy ($\Delta E_{ani}$), Dzyaloshinskii-Moriya interaction energy ($\Delta E_{DM}$) as [7, 8];

$$\Delta E_{spin} = \Delta E_{ex} + \Delta E_{ani} + \Delta E_{DM} \qquad \ldots \qquad (1)$$

### 3.1 Isotropic Exchange interaction

Exchange interaction between spin magnetic moment $S_i$ and $S_j$ is given by Heisenberg exchange interaction energy term.

$$\Delta E_{ex} = - \sum_{i>j} J \vec{S_i} \cdot \vec{S_j} = - \sum_{i>j} J S^2 \cos(\theta) \qquad \ldots \qquad (2)$$

Where $\theta$ is the angle between the two spin vectors.

$Fe_3O_4$ has inverse spinel configuration where $Fe^{+3}$ ions reside in the tetrahedral sites and both $Fe^{+2}$ and $Fe^{+3}$ ions reside at the octahedral sites. The overall magnetic orientations between $T_d$ and $O_h$ sites are opposite to each other and shows ferrimagnetic response. There are two different magnetic exchange mechanisms operative in this system. One is superexchange interaction which is mediated by O atoms between $Fe^{+3}$ at Td and $Fe^{+2}$ in Oh and a second double exchange interaction between $Fe^{+2}$ and $Fe^{+3}$ in the same octahedral sites. <u>Generally, superexchange interaction is dominating and responsible for the overall ferrimagnetic ordering in the system</u>.

Therefore, considering the dominant exchange interaction between $T_d$ and $O_h$ sites, the exchange interaction energy cost due to $T_d$ canting for one-unit cell of $Fe_3O_4$ can be written as;

$$\Delta E_{ex} = 8JS^2 Cos(\theta) \quad \ldots \quad (3)$$

Where θ is the canting angle.

### 3.2 Anisotropy Energy

The anisotropy energy cost for cubic system is given by the formula

$$\Delta E \approx K_u Sin^2(\theta) \quad \ldots \quad (4)$$

Where θ is the angle with respect to the easy axis. In case of $Fe_3O_4$ the easy axis is <111> and [100] below the Verwey transition temperature. The anisotropy energy cost associated with the spin canting for $T_d$ site is given below;

$$\Delta E_{ani} = K(5 - Cos(2\theta)) \quad \ldots \quad (5)$$

### 3.3 DM interaction energy

The DM interaction is given by the formula.

$$\Delta E_{DM} = -\sum_{i>j} D_{ij} \vec{S_I} \times \vec{S_J} \quad \ldots \quad (6)$$

Where D is the DM interaction coefficient and is perpendicular to the isotropic exchange interaction term. The final expression for $T_d$ spin canting for $Fe_3O_4$ reduces to,

$$\Delta E_{DM} = 8 DS^2 Sin(\theta) \quad \ldots \quad (7)$$

### 3.4 Total spin Hamiltonian Energy

Thus summing all the energy contributions for $Fe_3O_4$ the spin Hamiltonian can be written as:

$$\Delta E_{total} = 8JS^2 Cos(\theta) + K(5 - Cos(2\theta)) + 8DS^2 Sin(\theta) \qquad \ldots (8)$$

Additional energy cost arising from the spin canting as calculated by DFT is fitted with the above components of the spin Hamiltonian to extract various contributions to the overall energy cost.

**Table 3**

| | Fitted parameter from Spin Hamiltonian | | |
|---|---|---|---|
| | J<br>Isotropic Exchange<br>(meV) | K<br>Single Ion Anisotropy<br>(meV) | $D_z$<br>Dzyaloshinskii-Moriya Interaction<br>(meV) |
| $Fe_3O_4$ | - 0.39 | 1.15 | 0.69 |
| $CoFe_2O_4$ | - 0.08 | 0.11 | 0.05 |

**References**

[1] D. S. Negi, B. Loukya, K. Ramasamy, A. Gupta, and R. Datta, Appl. Phys. Lett. **106**, 182402 (2015).

[2] P. Blaha, K. Schwarz, G. K. H. Madsen, D. Kvasnicka, and J. Luitz, WIEN2k: *An Augmented Plane Wave + LocalOrbitals Program for Calculating Crystal Properties* (Karlheinz Schwarz, Techn. Universität Wien, Austria, 2001).

[3] H. T. Jeng, G. Y. Guo, and D. J. Huang, Phys. Rev. Let. **93**, 156403 (2004).

[4] N. M. Caffrey, D. Fritsch, T. Archer, S. Sanvito, and C. Ederer, Phys. Rev. B **87**, 024419 (2013).

[5] K. L. kryka, J. A. Borchers, R. A. Booth, Y. Ijiri, K. Hasz, J. J. Rhyne, and S. A. Majetich, Phys. Rev. Lett **113**, 147203 (2014).

[6] K. Hasz, Y. Ijiri, K. L. Krycka, J. A. Borchers, R. A. Booth, S. Oberdick, and S. A. Majetich, Phys. Rev. B **90,** 180405(R) (2014).

[7] P. Liu, S. Khmelevskyi, B. Kim, M. Marsman, D. Li. X. Q. Chen, D. D. Sharma, G. Kresse, and C. Franchini, Phys. Rev. B **92**, 054428 (2015).

[8] C. Weingart, N. Spaldin, and E. Bousquet, Phys. Rev. B **86**, 094413 (2012).